\documentclass[sigconf]{acmart}

\AtBeginDocument{%
  }

\setcopyright{acmlicensed}
\copyrightyear{2018}
\acmYear{2018}
\acmDOI{XXXXXXX.XXXXXXX}
\acmConference[Conference acronym 'XX]{Make sure to enter the correct
  conference title from your rights confirmation email}{June 03--05,
  2018}{Woodstock, NY}
\acmISBN{978-1-4503-XXXX-X/2018/06}

\usepackage{multirow}
\usepackage{threeparttable}
\usepackage{booktabs}
\usepackage{makecell}
\usepackage[most]{tcolorbox}
\usepackage{xcolor}

\usepackage{algorithm}
\usepackage{algorithmic}

\tcbset{
    colback=gray!5,
    colframe=black,
    boxrule=0.8pt,
    arc=4pt,
    left=6pt,
    right=6pt,
    top=6pt,
    bottom=6pt
}

\definecolor{ashgray}{RGB}{240,240,240}
\definecolor{lightbluebox}{RGB}{220,235,247}

\newtcolorbox{promptbox}{
    colback=ashgray,
    colframe=black,
    boxrule=0.5pt,
    arc=3pt,
    left=6pt,
    right=6pt,
    top=6pt,
    bottom=6pt
}

\begin{document}

\title{Pen-Strategist: A Reasoning Framework for Penetration Testing Strategy Formation and Analysis}





\author{Yasod Ginige*, Pasindu Marasinghe*, Sajal Jain\textsuperscript{\dag}, Suranga Seneviratne*\\
*\textit{The University of Sydney, NSW, Australia}\\
\textsuperscript{\dag}\textit{Catharsis.net.au, Australia}\\
Email: \{yasod.ginige, pasindu.marasinghe, suranga.seneviratne\}@sydney.edu.au, sajal@catharsis.net.au}

\renewcommand{\shortauthors}{Trovato et al.}

\begin{abstract}
  Cyber threats are rapidly increasing, expanding their impact from large-scale enterprises to government services and individual users, making robust security systems increasingly essential. However, a significant shortage of skilled cybersecurity professionals exacerbates this challenge. While recent research has explored automating tasks such as penetration testing using LLM-based agents, existing frameworks often perform poorly due to limited capability in strategy formulation, domain-specific reasoning, and accurate action and tool selection. To overcome these limitations, we propose \textit{Pen-Strategist} framework, consisting of a novel domain-specific reasoning model that derives pentesting strategies via logical reasoning and a classifier that converts the strategies into actionable steps. First, we construct a reasoning dataset containing logical explanations for both strategy derivation and step selection in pentesting scenarios. We then fine-tune a Qwen-3-14B model for strategy generation using reinforcement learning. Evaluation on the test split of the dataset demonstrates a 87\% improvement in strategy derivation performance compared to the baseline. Furthermore, we integrate the fine-tuned Pen-Strategist model into existing automated pentesting frameworks, such as PentestGPT, and evaluate its performance on vulnerable machines, achieving a 47.5\% improvement in subtask completion while surpassing the baseline GPT-5. Further experiments on the CTFKnow benchmark show an 18\% performance gain over the base model. For step prediction, we train a semantic-based CNN classifier, which outperforms commercial LLMs by 28\% and enhances execution stability. Finally, we conduct a user study to qualitatively assess the generated strategies, and Pen-Strategist demonstrates superior performance compared to the Claude-4.6-Sonnet.
\end{abstract}



\received{20 February 2007}
\received[revised]{12 March 2009}
\received[accepted]{5 June 2009}

\maketitle

\section{Introduction}

Cyber incidents and attacks are increasing rapidly worldwide. While attackers have traditionally focused on large enterprises, they now also target small and medium-sized enterprises (SMEs), public sector organizations, and critical services such as hospitals and emergency response systems. This expanded threat landscape requires systems of all sizes to maintain a robust and continuously evolving security posture. Penetration testing (pentesting) and vulnerability and threat assessment are standard practices for securing software and networked systems. Pentesting simulates real-world attacks to uncover exploitable weaknesses, while vulnerability and threat assessment systematically identifies, analyzes, and prioritizes risks to enable proactive mitigation. Their frequency is often driven by regulatory requirements that vary across industries and government sectors~\cite{fedramp_2024}. However, these processes are time-consuming and resource-intensive, and the cybersecurity workforce cannot keep pace with demand due to the persistent skills shortage~\cite{al2018study}. Consequently, automating these processes has become essential.


Recent studies have explored the use of LLMs and LLM-based agents to automate security tasks such as penetration testing~\cite{deng2024pentestgpt, ginige2025autopentester, shen2025pentestagent}. However, these approaches face several fundamental limitations: {\bf (i)} they often fail to select effective strategies across different stages of the pentesting process, leading to poor vulnerability exploitation and limited attack surface coverage, and consequently require expert human guidance, undermining full automation~\cite{deng2024pentestgpt, AutoAttacker, kong2025vulnbot}; {\bf (ii)} they rely on external commercial LLM APIs, which necessitate transmitting sensitive information such as network structure and potential vulnerabilities to third-party servers, raising security and privacy concerns, while also introducing non-trivial operational costs and limiting deployment in environments requiring strict data locality or low-latency, tightly coupled command execution~\cite{deng2024pentestgpt, shen2025pentestagent, ginige2025autopentester}; and {\bf (iii)} they exhibit unstable execution behavior, including hallucinations and invocation of incorrect or unavailable tools due to insufficient guidance in tool selection and orchestration~\cite{shen2025pentestagent, kong2025vulnbot}, during multi-step attack execution.

To this end, we propose Pen-Strategist, a framework that consists of two models; (i) Strategy model: a domain-specific reasoning model to logically derive strategies for pentesting stages by evaluating prior findings, and (ii) Step model: a step classifier that converts generated strategies into actionable steps and tool selections. The two models can be used locally to generate strategies for Penetration testing tasks, while protecting data privacy. Furthermore, they can be integrated with agentic frameworks such as PentestGPT~\cite{deng2024pentestgpt} for automated pentesting to achieve better results. First, we construct a novel reasoning-centric dataset tailored for penetration testing, comprising structured inputs along with logical explanations for both strategy derivation and step-level decision making. This dataset enables logically consistent reasoning while jointly supporting action and tool prediction, thereby reducing hallucinations and ensuring adherence to constraints in the execution environment. Subsequently, we train two models using this dataset. As the Strategy model, we finetune an open source Qwen-3-14B model, using Group Relative Policy Optimization (GRPO)~\cite{shao2024deepseekmath}, to improve its ability to generate logically grounded strategies across different stages of penetration testing. As the step model, we train a convolution-based classifier on top of frozen language model embeddings, which improves execution reliability and reduces invalid tool usage. We evaluate the Pen-Strategist framework on pentesting tasks and demonstrate significant performance improvements compared to commercial LLMs, driven by accurate strategy derivation and tool selection. Furthermore, we extend the evaluation to generalized red-teaming tasks, including CTF-style challenges, to assess the robustness beyond the training distribution. We also conduct ablation studies to analyze the contribution of different fine-tuning strategies and pipeline components. Finally, we conduct a survey with security professionals to qualitatively evaluate the practicality and effectiveness of the generated strategies.

More specifically, we make the following contributions.

\begin{itemize}
\item {We propose Pen-Strategist, a novel framework for pentesting strategy derivation and the next action prediction, consisting of two models: the Strategy model and the Step model.}
\item We create a new dataset to finetune LLMs for strategy derivation and step prediction in pentesting tasks. The dataset contains reasoning data for both strategy derivation and step prediction, covering 240 Hack-The-Box and VulnHub machines. To the best of our knowledge, the dataset is the first of its kind.
\item We fine-tune the Strategy model, built on an open-source LLM using reinforcement learning, achieving an 87\% performance improvement over the baseline. We also train a semantics-based dual-head CNN model (Step model) for action classification, which attains 82.8\% accuracy in step prediction and outperforms commercial LLMs.
\item We conduct extensive experiments by integrating Pen-Strategist into agentic pentesting frameworks (PentestGPT, AutoPentester, and VulnBot), achieving a 47.5\% improvement in subtask completion, in Hack-The-Box machines. We further evaluate generalizability on CTF challenges, where Pen-Strategist improves performance by 28\% over the base model.
\item Finally, we conduct a user survey among security professionals to qualitatively evaluate Pen-Strategist's strategies in comparison with GPT-5 and Claude-4.6-Sonnet. We show that our strategy analyzer stands as the first choice in 52.4\% of the cases, surpassing the other two models. 
\end{itemize}

We provide our dataset and code in our GitHub repository.\footnote{\textcolor{blue}{https://github.com/YasodGinige/Pentest-Strategist.git}}
\section{Related Work}\label{sec:related_work}

\subsection{Automated Penetration Testing}

Early approaches to automated penetration testing relied on traditional reinforcement learning methods. For instance, Hu et al.~\cite{hu2020automated} proposed a two-stage Deep Q-learning approach that first constructs an attack tree from network topology information and enumerates possible attack paths, then applies a Deep Q-Learning Network (DQN) to select the most easily exploitable path. However, their work focuses solely on recommending attack vectors rather than exploiting software vulnerabilities.


Recent advancements in large language models (LLMs), including GPT~\cite{brown2020language}, Claude~\cite{anthropic_2026}, and Gemini~\cite{team2023gemini}, have enabled new approaches to automating cybersecurity tasks through LLM-based agents. PentestGPT~\cite{deng2024pentestgpt} represents one of the first major efforts toward LLM-driven penetration testing, employing a summarizer-analyzer-generator pipeline. However, it remains semi-automated, requiring security professionals to manually execute strategies and provide feedback to guide the analyzer toward correct approaches. AutoAttacker~\cite{AutoAttacker} automated command generation and execution but was restricted to the Metasploit framework. The AutoPentester framework~\cite{ginige2025autopentester} addressed this limitation by supporting a broader range of tools; nevertheless, despite its multi-agent architecture, subtask completion rates on Hack-The-Box~\cite{hackthebox_2024} challenges remain low, primarily due to strategy identification failures. Similarly, PentestAgent~\cite{shen2025pentestagent} and VulnBot~\cite{kong2025vulnbot} employ multi-agent architectures with specialized agents for planning, execution, and summarization, yet they too suffer from low subtask completion rates. The fundamental limitation underlying these frameworks is that commercial LLMs lack specialized capabilities for analyzing penetration testing strategies, as they are not specifically fine-tuned for such tasks. Additionally, existing frameworks frequently fail during action execution due to incorrect or unavailable tool selection~\cite{deng2024pentestgpt, kong2025vulnbot}.

\subsection{Cybersecurity Datasets for LLM Evaluation}

Multiple datasets have been released to assess LLMs' capabilities in various cybersecurity tasks. For example, the ExCyTIn-Bench dataset~\cite{wu2025SecRL} is designed to evaluate LLM agents on their capabilities for cyber incident assessment, framed as multi-step question-answering over security logs. It is constructed from a controlled Azure environment, simulating eight real-world attack scenarios. The final dataset consists of 589 question–answer pairs; however, it does not provide any opportunity for strategic reasoning-based training, rather acts as a benchmark for existing LLMs. The CTFKnow dataset~\cite{CTFKnow} is designed for evaluating LLMs on automated vulnerability discovery and exploitation tasks, where models must identify and reason about potential attack vectors. It contains multiple choice questions; therefore cannot be used to fine-tune a model for strategy derivations. The CyberSecEval dataset~\cite{bhusal2024secure} is designed to evaluate LLMs on cybersecurity tasks such as vulnerability identification, exploit generation, and CVE score prediction. The dataset is constructed by combining real-world security benchmarks, including CVEs, CTF challenges, and code samples. However, none of the above datasets provides the platform to fine-tune a model for strategy analysis in the pentesting. CyberLLaMA~\cite{zhang2025cyberllama} introduces a named entity recognition (NER) framework aimed at extracting structured security information, such as threats, vulnerabilities, and malware, from unstructured text. It fine-tunes a LLaMA-3.2-3B model on a large, carefully curated corpus of cybersecurity articles annotated with 4,788 unique entities, improving sequence labeling performance.
Similarly, TrafficLLM~\cite{ginige2025trafficllm} explores the fine-tuning of large language models for network traffic analysis to detect potential security threats. However, neither of these approaches addresses the need for a reasoning-oriented model capable of deriving penetration testing strategies.

\subsection{Finetuning LLMs with Reinforcement Learning}

Reinforcement learning-based fine-tuning enables LLMs to develop advanced reasoning and strategic decision-making capabilities, essential for domains requiring sequential, goal-oriented problem-solving. Group Relative Policy Optimization (GRPO)~\cite{shao2024deepseekmath} demonstrated the scalability of this approach, establishing it as a viable method for creating specialized reasoning models. GRPO and its variants~\cite{qiu2025open, liu2026gdpo} extend preference-based finetuning such as Direct Preference Optimization (DPO)~\cite{rafailov2023direct} by sampling multiple candidate outputs for the same prompt and ranking them within a group, allowing the model to learn from relative quality signals rather than absolute rewards.


Generalized Direct Preference Optimization (GDPO)~\cite{liu2026gdpo} builds on direct preference learning by incorporating more flexible formulations of preference comparisons, enabling the model to better align with nuanced or structured feedback beyond simple pairwise rankings. It typically improves sample efficiency and reduces reliance on explicit reward modeling while maintaining stable optimization. Reinforcement Learning with Verifiable Rewards (RLVR)~\cite{wen2025rlvr} focuses on tasks where the correctness of the responses generated can be automatically checked, using deterministic reward signals derived from verifiers instead of learned reward models or human annotations. This approach reduces ambiguity in supervision and enables scalable training with reliable feedback. Together, these methods represent a shift toward more stable, scalable, and automation-friendly alignment techniques that reduce dependence on human labeling and complex reward modeling.

\begin{figure}[t]
    \centering
    \includegraphics[width=\linewidth]{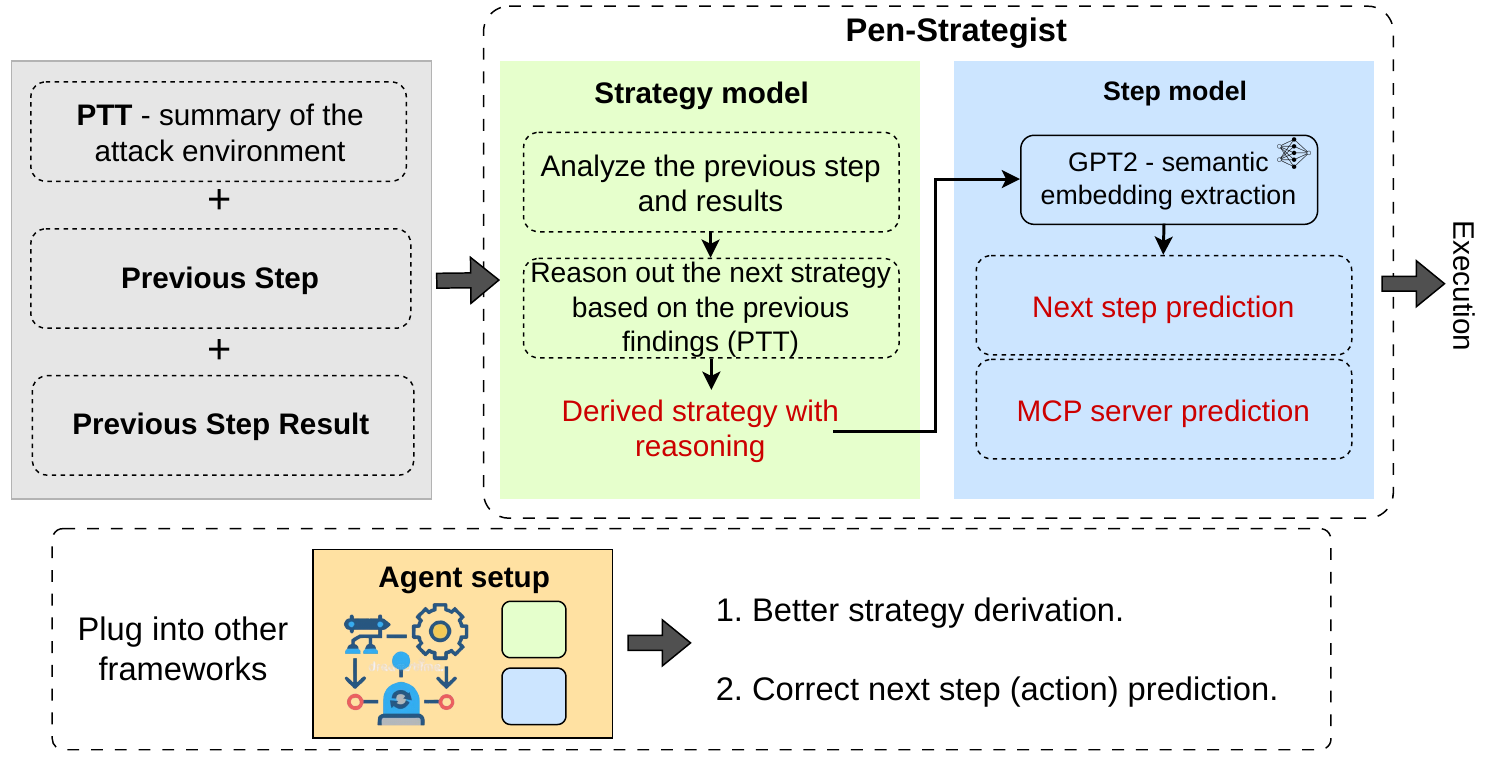}
    \caption{System overview of the Pen-Strategist. The framework is designed to be modular, allowing integration with other agent-based architectures.}
    \label{fig:system_overview}
\end{figure}

Fine-tuning LLMs using reinforcement learning, in particular GRPO, for reasoning has shown promising results beyond typical scenarios such as coding and mathematics. For instance,~\cite{dai2025qoq,qiu2025open} reported that RL improves clinical reasoning and diagnostic generation under data scarcity. Similarly, Dai et al.~\cite{dai2025legal} found that RL can be used to improve structured reasoning in legal applications. Similar improvements have been shown in finance for tasks such as credit assessment and risk pricing~\cite{liu2025fin}. However, to the best of our knowledge, no existing work has fine-tuned a reasoning model specifically for penetration testing tasks.


\section{System Overview}\label{subsec:system_overview}

\begin{figure*}
    \centering
    \includegraphics[width=\linewidth]{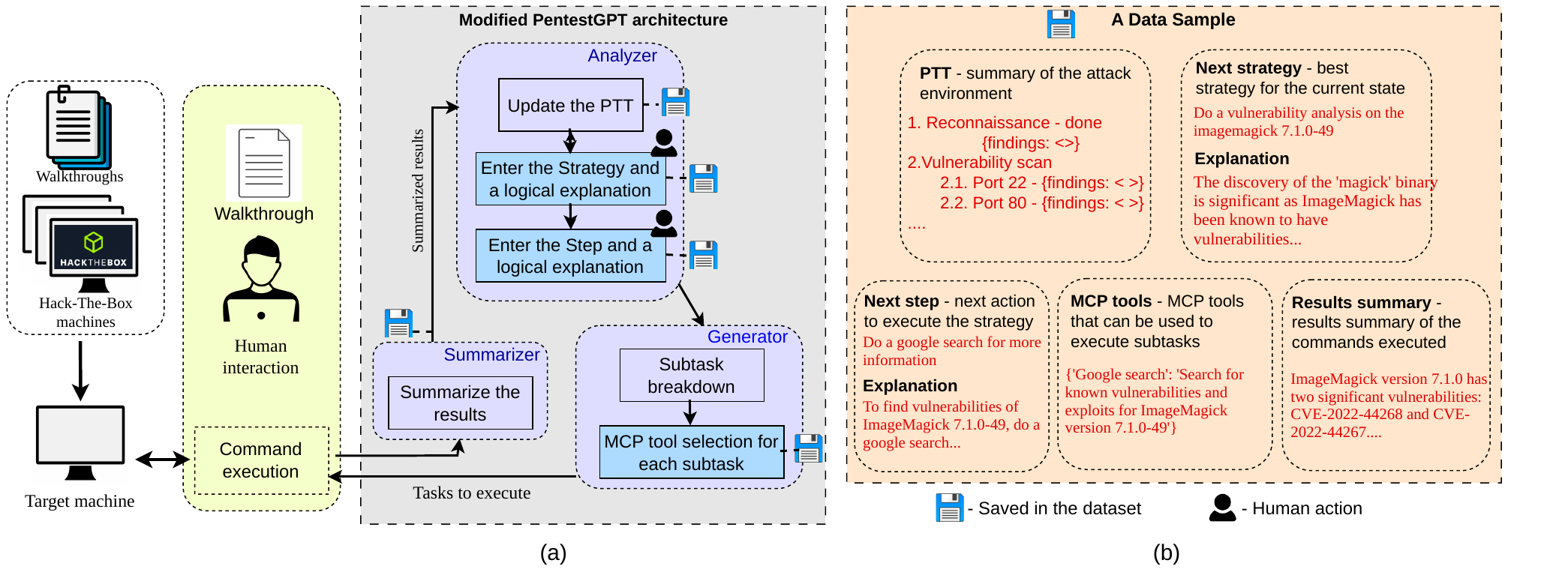}
    \caption{Dataset collection steps (a) and a sample data point (b).}
    \label{fig:data_collection}
\end{figure*}

As illustrated in Figure~\ref{fig:system_overview}, Pen-Strategist is a modular framework comprising two models for strategy generation and step (i.e., action) classification in the pentesting. The \textit{Strategy model} acts as a domain-specific reasoning component that derives logical strategies for different pentesting stages based on prior findings, while the \textit{Step model} functions as a classifier that translates these strategies into executable steps and tool selections, guiding the automated execution. Pen-Strategist is designed to integrate with existing agentic penetration testing frameworks (e.g., PentestGPT~\cite{deng2024pentestgpt}, AutoPentester~\cite{ginige2025autopentester}), improving overall task performance.

The Strategy model is fine-tuned to generate logically consistent strategies based on a summary of the target environment, including previously executed steps and their corresponding findings. To this end, we fine-tune a Qwen-3-14B model with Low-Rank Adaptation (LoRA)~\cite{hu2022lora}, as detailed in Section~\ref{subsec:model_train}.

The Step model is a multi-label classifier designed to predict the next action from a given strategy, along with the MCP servers required to execute it. The strategy is first encoded into a context-aware semantic embedding using a GPT-2 model. This embedding is then fed into a dual-head CNN, where one head predicts the next step (action) and the other identifies the relevant MCP servers. As a result, a follow-up LLM can be guided to decompose the main step into substeps to execute the strategy using the selected MCP servers. Having a separate step model reduces the hallucinations and execution failures due to incorrect tool usage. For instance, if the execution environment does not support the \textit{Gobuster} tool but does support \textit{Dirbuster}, only the \textit{Dirbuster} MCP tool will be included in the classification. This allows a downstream LLM to be explicitly guided to use Dirbuster, which would not be reliably ensured if it were simply asked to generate commands for the chosen steps—potentially leading to execution errors. 
 Furthermore, the Step Model gives access to widely used security MCP servers and enables individual pentesters to use their preferred subset of tools (from a large and diverse ecosystem ) with existing frameworks by simply finetuning the classifier. The MCP servers can also be replaced using recently developed Claude Skills~\cite{agent_skills_2025}, which will be discussed in the Discussion. Overall, the Pen-Strategist framework manages the full planning pipeline, from deriving a logical strategy to its high-level execution. The significance of these components is discussed in the ablation study (Section~\ref{subsec:results_setup_D}). 


\section{Methodology}\label{sec:methodology}

In this section, we discuss curation of the training dataset, followed by Pen-Strategist model training. 


\subsection{Dataset Curation}\label{subsec:dataset_collection}

Our dataset is collected to accomplish two tasks: fine-tuning and training models for (i) pentest strategy derivation through reasoning and (ii) predicting the next step to execute a given strategy. It contains data points collected using 240 HTB and VulnHub machines in total, under two main parts based on the data curation process: {\bf (i) Manual collection:} We collected data for 40 HTB machines using human supervision to input the strategy, next steps, and the logical reasoning for each decision using the write-ups (further explained in Section~\ref{subsubsec: manual_collection}). Here, the human annotator breaks into HTB machines following walkthroughs and records the steps with logical reasoning for each decision, (ii) {\bf Automated collection:} Since it is not practical to complete all the machines manually, and the fact that a large number of data points are required to fine-tune a reasoning model, we also designed an automated data curation workflow using Claude Code as detailed in Section~\ref{subsubsec: automatic_collection}. Our complete dataset contains 2,165 data points.

\subsubsection{\textbf{Manual data collection workflow}}\label{subsubsec: manual_collection}

As illustrated in Figure~\ref{fig:data_collection}-(a), in the manual data collection workflow, we employ a modified PentestGPT framework and systematically log the required data fields while iteratively following the penetration testing workflow based on the corresponding machine write-ups. PentestGPT is modified in two key ways. First, instead of Analyzer automatically generating the next optimal strategy and steps, a human annotator provides the strategy, action, and their associated logical explanations by consulting the write-ups. Second, during execution, we record a set of structured data points at each iteration. Specifically, we capture the \textit{PenTest Tree (PTT)}, which summarizes the current state of the target environment, including prior actions and their outcomes. We also log the \textit{Previous step} and \textit{Previous step result}, representing the most recent action and its observed outcome. Furthermore, the \textit{New strategy} and \textit{Strategy explanation} describe the next optimal strategy and its underlying rationale, derived from the PTT and preceding results. The \textit{New step} and \textit{Step explanation} detail the concrete action to be executed, including the tools or MCP servers involved and their usage. Finally, the \textit{Results} field records the outputs obtained from executing the proposed step. A detailed discussion about the data fields and formats is given in Section~\ref{subsubsec:dataset_fomalities}.

\subsubsection{\textbf{Automated data collection workflow}}\label{subsubsec: automatic_collection}

In the automated data collection workflow, we use Claude Code to extend the dataset using an additional 200 machines from HTB and VulnHub. Initially, we compile a list of target machines from both platforms and collect their corresponding write-up files. Next, we provide Claude Code with a subset of manually curated dataset samples along with their associated write-ups, enabling it to infer the desired data structure and formatting. Subsequently, guided by explicit instructions and formatting specifications, Claude Code is used to transform the information contained in the write-ups into the required dataset format. Importantly, this process is purely a data conversion and structuring task, where Claude Code extracts and reformats information from existing write-ups, rather than executing any live commands or performing autonomous penetration testing. As the source write-ups are already validated and the manually collected samples serve as reliable references, the generated dataset is consistently well-structured. To further ensure data quality, we perform a manual review of the entire dataset to validate formatting consistency, along with an additional inspection of 10\% randomly sampled entries to verify their content correctness against the original write-ups. This process did not reveal any significant errors, indicating the dataset's overall reliability. 



\subsubsection{\textbf{Dataset format}}\label{subsubsec:dataset_fomalities}

Figure~\ref{fig:data_collection}-(b) illustrates a sample data instance comprising: \textit{PTT}, which represents the pentesting process as an attack tree enriched with summarized findings; \textit{New strategy} and \textit{Strategy explanation}, which provide the ground truth strategy and its derivation for a given pentest instance; \textit{Next step} and \textit{Step explanation}, which specify the action to be taken along with its rationale; \textit{MCP tools}, which gives suitable tools to execute the selected step; and \textit{Results}, which summarize the tool outputs from executed commands. 

\begin{figure}[t]
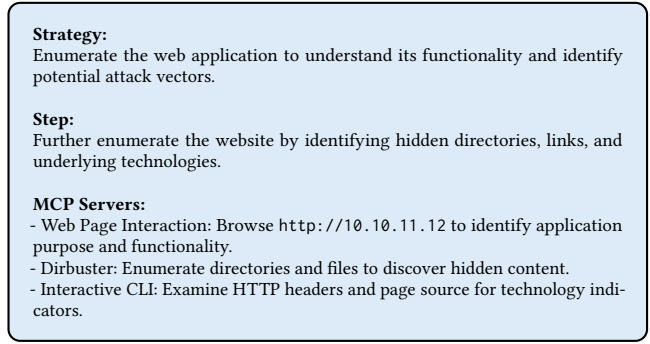

\centering
\footnotesize

\begin{tcolorbox}[
    colback=lightbluebox,
    colframe=black,
    width=\linewidth
]

\textbf{Strategy:} \\
Enumerate the web application to understand its functionality and identify potential attack vectors.\\

\textbf{Step:} \\
Further enumerate the website by identifying hidden directories, links, and underlying technologies.\\

\textbf{MCP Servers:}\\ 
- Web Page Interaction: Browse \texttt{http://10.10.11.12} to identify application purpose and functionality.\\
- Dirbuster: Enumerate directories and files to discover hidden content.\\
- Interactive CLI: Examine HTTP headers and page source for technology indicators.

\end{tcolorbox}

\caption{A sample strategy decomposition illustrating the use of MCP servers for web application enumeration.}
\label{fig:enumeration_example}

\end{figure}

The \textit{Next step} is restricted to a predefined set of high-level actions as listed in the first column of Table~\ref{tab:next-step}. This constraint helps stabilize the automated execution phase by reducing hallucinations and variability in command generation and tool usage. As illustrated in Figure~\ref{fig:enumeration_example}, when executing a strategy, the step model classifies it into one of these steps and predicts the MCP server to be used to execute the step. We define 11 \textit{MCP servers}\footnote{An MCP server is a standardized, secure bridge that connects AI models to external tools and data sources. Example security-specific MCP servers can be found here: https://github.com/cyproxio/mcp-for-security.} that commonly used in pentesting: Nmap, Metasploit, Netcat, Dirbuster, SQLmap, SMB client, Hydra, John-the-ripper, Google search, Interactive CLI, and Web page interaction. This facilitates training a model to accurately produce the New step and its explanation, while also selecting the appropriate MCP servers needed to execute the commands for that step.


\subsubsection{\textbf{Dataset summary}}\label{subsubsec:dataset_analysis}

\begin{figure}
    \centering
    \includegraphics[width=\linewidth]{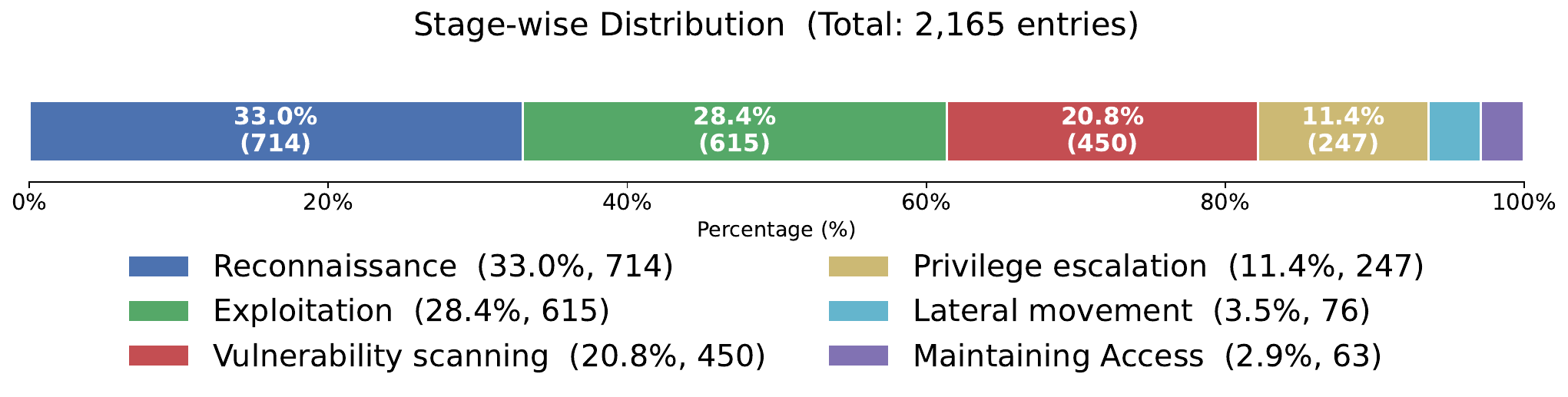}
    \caption{Distribution of the strategies in different stages of the pentesting process.}
    \label{fig:data_analysis}
\end{figure}

Next, we present summary statistics of the dataset. First, we examine the dataset by analyzing how strategies as classified by an LLM are distributed across pentesting stages (e.g., reconnaissance, vulnerability scanning, and others), as shown in Figure~\ref{fig:data_analysis}. In this stage-wise breakdown, reconnaissance (33.0\%) and exploitation (28.4\%) represent the largest portions of the data, followed by vulnerability scanning (20.8\%). Other stages, including privilege escalation, lateral movement, and maintaining access, are less frequently observed. 
This distribution is advantageous for model fine-tuning, as it reflects realistic operational patterns and allows the model to focus on the most critical and commonly occurring stages, particularly reconnaissance and exploitation. At the same time, the inclusion of 76 lateral movement and 63 maintaining access instances ensures that even these less frequent stages are adequately represented, enabling the model to learn a more complete spectrum of pentesting tasks. 

Table~\ref{tab:next-step} lists the distribution of the \textit{Next Steps} in the dataset. 
In particular, exploiting identified vulnerabilities (35.64\%) and further enumeration of services (20.26\%) appear most frequently, followed by activities such as analyzing findings and reporting-related steps. Other actions, such as domain enumeration and source code analysis, occur less frequently due to their lower prevalence in the HTB environment.

\begin{table}[t]
\footnotesize
\centering
\caption{Distribution of the items of \textit{Next Step}}
\label{tab:next-step}
\begin{tabular}{p{0.65\linewidth}rr}
\toprule
Next step & Count & \% \\
\midrule
Exploit the selected exploitations & 628 & 35.64 \\
Enumerate further on the X service to find software versions, hidden directories and file. & 357 & 20.26 \\
Explore the suspicious files, commands and create a summary of the findings. & 176 & 9.99 \\
End task and ask permission to generate the report & 174 & 9.88 \\
Further enumerate the website (hidden directories, links, software) & 165 & 9.36 \\
Do a Google search for more information & 108 & 6.13 \\
Analyze the outcomes of the previous step and find an attack path & 93 & 5.28 \\
Enumerate the domain & 31 & 1.76 \\
Explore the source code for vulnerabilities & 29 & 1.65 \\
\bottomrule
\end{tabular}
\end{table}


\subsection{Pen-Strategist Model Training}\label{subsec:model_train}

As mentioned before, the Pen-Strategist framework consists of two key components: \textit{strategy prediction} and \textit{step prediction}, capturing the logical (what to do) and methodological (how to do it) aspects of pentesting, respectively. To this end, we develop two specialized models. The strategy model derives the most appropriate next strategy based on prior steps and observed findings, while the step model predicts the next action along with the corresponding MCP servers required for execution. We fine-tune a Qwen3-14B~\cite{Qwen3_14B} model with LoRA~\cite{hu2022lora} on a single H200 GPU for strategy prediction, and train a semantics-based dual-head CNN model for step classification. 

\subsubsection{\textbf{Strategy Model}}\label{subsubsec:strategy_model}

\begin{figure}
    \centering
    \includegraphics[width=\linewidth]{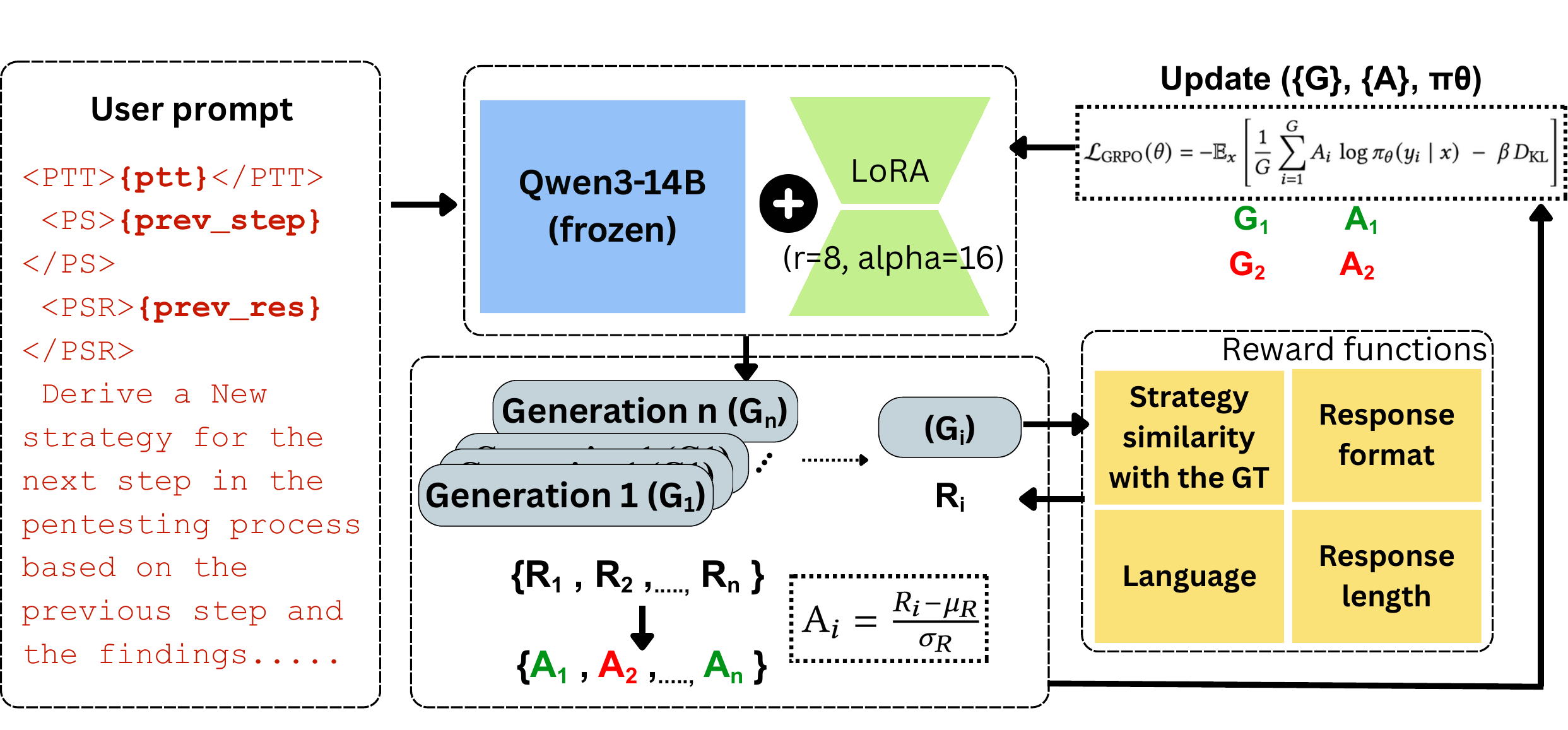}
    \caption{Strategy model training steps.}
    \label{fig:grpo}
\end{figure}

As illustrated in Figure~\ref{fig:grpo}, we fine-tune the Qwen3-14B model using LoRA with GRPO~\cite{shao2024deepseekmath}, where only the LoRA parameters are updated during training. The objective is to enable the model to infer the next strategy in a pentesting workflow by emulating human reasoning, analyzing prior findings to determine subsequent actions. To provide sufficient context, the user prompt includes the PTT, which summarizes the attack environment up to the current stage and includes executed steps and their outcomes, along with the immediate previous step and its result. Based on this information, the model is guided to generate the next strategy through logical reasoning followed by a concise explanation. The detailed prompts used are provided in Appendix~\ref{app:prompts}. \\

\noindent{\textbf{GRPO fine-tuning}}

In GRPO, the model produces multiple candidate outputs ($y_i$) for each input prompt instead of a single response; in our setup, four responses are sampled from the current policy to form a comparison group. These candidates are evaluated using a set of reward functions, and the higher-quality outputs are selected based on their relative reward scores within the group. Learning is then driven by optimizing the model to increase the likelihood of these better-performing responses. This relative comparison-based optimization encourages the model to favor higher-quality outputs while improving training stability and reducing variance. The reward functions used are as follows: \\ \vspace{-3mm}

\noindent\textbf{i) Semantic and logical similarity reward ($R_s$):}
We quantify the semantic and logical alignment between the generated \emph{New strategy} and \emph{Strategy explanation} and their ground-truth counterparts using the G-Eval~\cite{geval} - an LLM as a judge framework, which employs an LLM to evaluate two texts under a fixed rubric. We use GPT-4o as the evaluator with four criteria ($GEval_k$) based evaluation in the GEval framework; logical alignment with the ground-truth rationale, reference to similar evidence and primary task, consistency of the final decision for the given context, and usage of similar tools and techniques. Each criterion is independently scored by the evaluator, and the final similarity reward is computed by averaging the scores given to each criterion, as denoted by \( R_{\text{s}}(g,y) = \frac{1}{4} \sum_{k=1}^{4} \mathrm{GEval}_k(g,y) \).


This reward captures the overall logical and semantic agreement between the generated strategy (y) content and the ground truth (g). The prompt used is provided in Figure~\ref{fig:reward_strategy} in Appendix~\ref{app:prompts}.\\ 

\vspace{-2mm}
\noindent{ii) \textbf{Pattern reward ($R_p$)}} forces the model to generate the output in the following pattern.
\textit{<think> logical derivation </think> New strategy <explanation> Strategy explanation </explanation>}. This is a hard reward where the exact match gets a 1, otherwise 0.\\

\vspace{-2mm}
\noindent{iii) \textbf{Generation length reward ($R_l$)}} the model to keep the generation length below the maximum token count 1,024. Otherwise, the reasoning tends to explode as training progresses. The reward function is defined according to the Equations~\ref{eq:delta} and~\ref{eq:reward}.

\begin{align}
\delta &= \max\!\left(0,\; \text{generation length} - \texttt{max\_token\_length}\right) \label{eq:delta} \\
R_{l} &= 1 - \frac{1}{2}\left(\frac{\delta}{\texttt{max\_token\_length}}\right) \label{eq:reward}
\end{align}
\\
\noindent{iv) \textbf{Language reward ($R_e$)}} forces the model to stick to the English language without mixing with other languages during the reasoning. This ensures the generated rationale is directly applicable to the similarity reward without any complexities. The language reward is defined as 1.0 if the output is in English, 0.0 if it is in any other language, and -1.0 if the output is empty.\\ 

Finally, we sum up all reward components into a single total reward, denoted as $R_{\text{total}}$. Since four strategies are generated for each prompt, we obtain four corresponding $R_{\text{total}}$ values, each reflecting the quality of the respective strategy. \( \mathcal{R}_{\text{total}} = \{ R_{\text{total}}(y_1), R_{\text{total}}(y_2),\\
R_{\text{total}}(y_3), R_{\text{total}}(y_4) \} \). 

\begin{align}
A_i &= \frac{R_{\text{total}}(y_i) - \overline{\mathcal{R}}_{\text{total}}}{\sigma{(\mathcal{R}_{\text{total}})} + \epsilon}, \quad i \in \{1,2,3,4\}\label{eq:advantage}
\end{align}

To encourage higher-quality strategies while discouraging weaker ones, GRPO assigns each sampled completion an advantage score based on its relative performance within the group using the Equation~\ref{eq:advantage}. For a given prompt, we sample $N=4$ candidate completions $\{y_i\}_{i=1}^{N}$, each of which is evaluated using a set of reward functions. The advantage $A_i$ is computed as the normalized deviation of the total reward of $y_i$ from the group mean, where positive values indicate above-average performance and negative values indicate below-average quality. 

\begin{align}
\mathcal{L}_{\text{GRPO}}(\theta)
=
- \mathbb{E}_{x}\left[
\frac{1}{N}\sum_{i=1}^{N}
A_i \, \log \pi_{\theta}(y_i \mid x)
-
\beta \, D_{\mathrm{KL}}\!\left(\pi_{\theta} \,\|\, \pi_{\text{ref}}\right)
\right]\label{eq:grpo_loss}
\end{align}

The policy is then optimized using an advantage-weighted objective, as defined in Equation~\ref{eq:grpo_loss}, which increases the likelihood of higher-advantage completions while reducing that of lower-advantage ones. In parallel, a Kullback--Leibler (KL) divergence~\cite{kullback1951kullback} penalty is applied to constrain the learned policy $\pi_{\theta}$ from deviating excessively from a fixed reference policy $\pi_{\text{ref}}$, which corresponds to the base pretrained model. This regularization term improves training stability and prevents overfitting to high-reward samples by maintaining proximity to the reference distribution.





We set the optimization learning rate to $5\time10^{-5}$, a common value for finetuning LLMs. The model was trained with the AdamW optimizer and a weight decay of 0.01 to regularize the parameters. A linear learning rate schedule was applied, in which the learning rate was gradually increased over the first 10\% of training steps (warmup ratio of 0.1) and then linearly decreased for the rest of the training.

\subsubsection{\textbf{Step Model}}\label{subsubsec:step_model}

We train a dual-task architecture consisting of a frozen GPT-2 encoder and two independent convolution-based classification heads to predict (i) the next step class and (ii) the set of MCP servers, conditioned on the \emph{New strategy} and its \emph{Strategy explanation}. The objective is to discourage invalid or unavailable tool selection and reduce overly generic or unfocused tool usage, thereby improving the stability and consistency of strategy execution.

Given the input formed by concatenating the strategy and its explanation, we extract contextual token-level embeddings using a pretrained GPT-2 model. Since GPT-2 does not provide a dedicated classification token, we retain full sequence representations and do not perform CLS pooling. The encoder is kept frozen and used solely as a feature extractor. On top of these embeddings, we apply two separate convolutional encoders with multiple kernel sizes to capture local semantic patterns. Each encoder performs temporal convolution followed by global max pooling to produce a fixed-dimensional representation. One representation is used for step classification, and the other for MCP server prediction.

\begin{algorithm}[b]
\caption{Dual-Task Training for step and MCP prediction with Frozen GPT-2 and dual head CNN model.}
\label{alg:dual_task_training}
\begin{algorithmic}[1]

\REQUIRE Dataset $\mathcal{D}$ with inputs $x$, step labels $y^{(s)}$, and MCP labels $y^{(m)}$
\REQUIRE Frozen GPT-2 encoder $f_{\theta}$, dual CNN heads $g_{\phi_s}$ and $g_{\phi_m}$
\REQUIRE Learning rate $\eta_t$, batch size $B$, epochs $E$, weight decay $wd$, and loss weights $\lambda_s, \lambda_m$

\STATE Initialize $\phi_s, \phi_m$; freeze $\theta$

\FOR{epoch $= 1$ to $E$}
    \FOR{each mini-batch $\{(x_i, y^{(s)}_i, y^{(m)}_i)\}_{i=1}^{B}$}
        
        \STATE $x_i \leftarrow \text{concat}(\text{New strategy}, \text{Strategy explanation})$
        
        \STATE $H_i \leftarrow f_{\theta}(x_i)$ \hfill // contextual embeddings
        
        \STATE $z^{(s)}_i \leftarrow g_{\phi_s}(H_i)$
        \STATE $z^{(m)}_i \leftarrow g_{\phi_m}(H_i)$
        
        \STATE $\mathcal{L}_s \leftarrow \text{CrossEntropy}(z^{(s)}_i, y^{(s)}_i)$
        \STATE $\mathcal{L}_m \leftarrow \text{BCEWithLogits}(z^{(m)}_i, y^{(m)}_i)$
        
        \STATE $\mathcal{L} \leftarrow \lambda_s \mathcal{L}_s + \lambda_m \mathcal{L}_m$
        
        \STATE Compute gradient: \( g \leftarrow \nabla_{\phi_s, \phi_m} \mathcal{L}_{\mathrm{total}} \)
        
        \STATE Update \( \phi_s, \phi_m \) using AdamW optimizer with \( \eta_t \) and \( wd \)

    \ENDFOR
\ENDFOR

\RETURN $\phi_s, \phi_m$

\end{algorithmic}
\end{algorithm}

We formulate the learning problem as a two-task multi-label classification setting, where a shared representation is used to jointly predict (i) the next step label and (ii) the MCP server set. For each training instance, we compute a cross-entropy loss for step prediction and a binary cross-entropy loss for MCP prediction using the ground-truth labels $y_{\mathrm{step}} \in A$ and multi-hot vector $y_{\mathrm{mcp}} \in \{0,1\}^{|M|}$, respectively:

{\footnotesize
\[
\mathcal{L}_{\mathrm{step}} = -\log p^{(\mathrm{step})}_{y_{\mathrm{step}}}
\]
}

{\footnotesize
\[
\mathcal{L}_{\mathrm{mcp}} =
- \sum_{j=1}^{|M|}
\Big[
y_{\mathrm{mcp},j} \log \sigma(z^{(\mathrm{mcp})}_j)
+ (1 - y_{\mathrm{mcp},j}) \log (1 - \sigma(z^{(\mathrm{mcp})}_j))
\Big]
\]
}

where $z^{(\mathrm{mcp})}_j$ denotes the logit corresponding to the $j$-th MCP class and $\sigma(\cdot)$ is the sigmoid activation. The final objective is a weighted combination of both losses, calculated as \( \mathcal{L} = \lambda_{\mathrm{step}} \mathcal{L}_{\mathrm{step}} + \lambda_{\mathrm{mcp}} \mathcal{L}_{\mathrm{mcp}} \).

where we set $\lambda_{\mathrm{step}} = 1$ and $\lambda_{\mathrm{mcp}} = 1.5$ based on empirical performance. Optimization is performed using AdamW with a linear warmup followed by linear decay, where $\eta_t$ denotes the time-dependent learning rate and $wd$ is the weight decay coefficient. The overall training procedure is summarized in Algorithm~\ref{alg:dual_task_training}. 


\section{Experimental Setup}\label{sec:experiments}

We divide the experiments into several setups for easier analysis.

\subsection{Test Set Evaluation} \label{subsec:exp_setup_A}

We evaluate the fine-tuned strategy and step models using the held-out test set of the dataset. It contains 10 machines from the manual collection and 30 from the automated collection. The goal is to measure how well the fine-tuned model performs in strategy derivation and step prediction compared to other commercial models such as GPT, Gemini, and Claude (The exact model versions are listed in the result Table~\ref{tab:model_strategy_explanation}). For strategy derivation, we use two metrics: final strategy similarity and the explanation similarities measured by G-Eval scores, following the same criteria as the similarity reward (Section~\ref{subsubsec:strategy_model}).

For step prediction and MCP server prediction, we use accuracy and Micro F1 score as evaluation metrics, respectively. Accuracy measures the model’s ability to correctly predict the exact next step as the ground truth. In contrast, MCP server prediction is a multi-label task, as a single step may involve multiple servers. Therefore, in addition to accuracy, which captures the correctness of the entire predicted set, we use the Micro F1 score to evaluate element-wise prediction performance. The F1 score is computed at the per-sample level and then averaged across all samples to obtain the final Micro F1 score. For each sample $n$, the true positives ($TP_n$), false positives ($FP_n$), and false negatives ($FN_n$) are computed as:

{\footnotesize
\[
TP_n=\sum_{j=1}^{K}\hat{y}_{n,j}\,y_{n,j},
\qquad
FP_n=\sum_{j=1}^{K}\hat{y}_{n,j}\,(1-y_{n,j}),
\qquad
FN_n=\sum_{j=1}^{K}(1-\hat{y}_{n,j})\,y_{n,j}.
\]
}

Here, \(N\) denotes the total number of samples in the evaluation set and \(K\) denotes the number of MCP server labels in the multi-label prediction space. For each sample \(n\in\{1,\dots,N\}\) and label \(j\in\{1,\dots,K\}\), the ground-truth indicator \(y_{n,j}\in\{0,1\}\) specifies whether MCP label \(j\) is actually present, while the predicted indicator \(\hat{y}_{n,j}\in\{0,1\}\) specifies whether the model predicts that label.
The corresponding results are presented in Section~\ref{subsec:results_dataset_evaluation}.

\subsection{Evaluation by Integrating to Existing Frameworks}\label{subsec:setup B}

In this experiment, we replace the strategy analysis/planning module of existing pentesting frameworks with the fine-tuned strategy model. Specifically, we evaluate this integration within the PentestGPT~\cite{deng2024pentestgpt}, AutoPentester~\cite{ginige2025autopentester}, and VulnBot~\cite{kong2025vulnbot} frameworks. As the evaluation benchmark, we use six HTB machines (Sau, Pilgrimage, Authority, Jupiter, Jarvis, and Bank), none of which were included in the training data of the fine-tuned model. The objective is to assess the performance improvement of these frameworks when employing the fine-tuned LLM in place of commercial LLMs within the strategy analysis component. For comparison, GPT-5 is used as the baseline commercial LLM under the default settings of each framework. As the evaluation metric, we use the subtask completion rate, a widely adopted metric in agent evaluations on pentesting tasks~\cite{deng2024pentestgpt,ginige2025autopentester, shen2025pentestagent}, where each machine's attack vector is divided into subtasks. We calculate the percentage of subtasks completed out of the total number of subtasks as the completion rate. For each machine, we conduct three runs and report the average subtask completion rate to minimize the variability of results.

Recent updates of Claud Code demonstrate enhanced skills in task automation. Therefore, we evaluate the Claude Code in pentesting tasks by giving the six HTB machines to solve autonomously. The corresponding results are presented in Section~\ref{subsec:results_setup_B}.

\subsection{Cross-Task Generalization}\label{subsec:setup_c}

To assess generalizability, we evaluate Pen-Strategist on additional security tasks that require logical reasoning, such as CTF challenges. In particular, we use the PicoCTF challenges~\cite{deng2024pentestgpt} and the CTFKnow~\cite{CTFKnow} dataset to determine whether fine-tuning improves performance in task completion. 

For the PicoCTF evaluation, we adapt the experiment setup in PentestGPT~\cite{deng2024pentestgpt}, which contains 20 challenges. PentestGPT uses GPT-5 as the backend for the strategy analyzer agent to solve the challenges; we replace it with the base Qwen-3-14B model and the fine-tuned strategy model, enabling us to assess the impact of fine-tuning on CTF problem-solving performance, which is measured using the challenge completion rate. For each challenge, we perform five independent runs and record the number of successful completions. 

In the CTFKnow dataset, we run the benchmark experiments using the fine-tuned strategy model directly. The dataset consists of multiple-choice questions based on security incidents, where the agent is required to analyze each scenario and select the most appropriate answer. We evaluate the task success rate, the original metric used in the benchmark.
The corresponding results are presented in Section~\ref{subsec:results_setup_c}.

\subsection{Ablation Study}\label{subsec:setup_d}

We conduct an ablation study using the AutoPentester~\cite{ginige2025autopentester} and integrate the fine-tuned strategy model as the backend of its analyzer agent. The Step model is used to predict the next step and tools that guide the command generator in the execution phase. Unlike PentestGPT, which requires manual command execution, AutoPentester is a fully automated framework. Therefore, it enables us to assess the effectiveness of the Step model by measuring reductions in command execution failures in an automated setting. 


For evaluation, we use three HTB machines—Sau, Authority, and Jarvis—as the test bench. We compare three configurations: (i) the original AutoPentester, (ii) AutoPentester with the fine-tuned strategy model, and (iii) AutoPentester with both the fine-tuned strategy model and the additional step prediction stage. This setup allows us to analyze the contribution of individual components, namely the strategy model and the step model, to overall system performance. The evaluation metric used is subtask completion rate. The corresponding results are presented in Section~\ref{subsec:results_setup_D}.

\subsection{Human Expert Evaluation}\label{subsec:setup_e}
We conducted a user study with 12 security experts to qualitatively evaluate the reasoning quality and practical usefulness of the generated pentesting strategies. The study was approved by our institution’s Human Research Ethics Committee, and the participants were recruited based on their expressed interest in a LinkedIn post promoting the survey. We used 15 pentesting scenarios, each outlining the sequence of steps performed and the corresponding findings describing a specific stage of an ongoing penetration testing process. These scenarios were organized into three sets, each consisting of five scenarios. For each scenario, we give three strategy outputs to rank generated by different LLMs: our fine-tuned Strategy model, GPT-5, and Claude-4.6-Sonnet. 
To avoid bias, model identities were anonymized. A sample scenario is given in Figure~\ref{fig:survey_sample} in the Appendix~\ref{app:prompts}.

Each participant was presented with one set containing five scenarios. For each scenario, they were asked to rank the strategies generated by three models based on logical correctness and alignment with the given task, assigning a rank of 1 to the best-performing strategy and 3 to the least effective. In addition, participants reported their confidence in each ranking decision using a three-level scale (very confident, somewhat confident, not at all confident) to account for potential ambiguity in judgment. Finally, free-text feedback was collected to better understand the reasoning behind participant rankings, and participants were also encouraged to provide qualitative comments on the strengths and weaknesses of the generated strategies.

 As evaluation metrics, we report: (i) the percentage of a particular model being selected as the first choice, and (ii) Kendall’s W statistic~\cite{lewis1971kendall} to assess the level of agreement among participants' rankings. To incorporate participant confidence into the evaluation, we filter a subset of scenarios where the participant was highly confident and recalculate the metrics. The corresponding results are presented in Section~\ref{subsec:results_setup_E}

\section{Results}\label{sec:results}

In this section, we report results from the experiments described in Section~\ref{sec:experiments}.

\begin{table}[t]
\centering
\caption{Model performance on Strategy and Explanation. Here, we use the GEval score as the evaluation criterion.}
\label{tab:model_strategy_explanation}
\begin{tabular}{lcc}
\hline
\textbf{Model Name} & \textbf{Strategy} & \textbf{Explanation} \\
\hline
Claude 3 Haiku         & 0.54 & 0.58 \\
Claude 4.5 Sonnet      & 0.65 & \textbf{0.72} \\
Gemini 2.0 Flash       & 0.40 & 0.53 \\
Gemini 2.5 Flash       & 0.45 & 0.56 \\
GPT-3.5-turbo          & 0.36 & 0.47 \\
GPT 4.1                & 0.60 & 0.66 \\
GPT-4o-mini            & 0.52 & 0.64 \\
GPT-5                  & 0.62 & 0.63 \\
LLaMA-3.1-8B           & 0.40 & 0.51 \\
\hline
Qwen-3-14B             & 0.39 & 0.45 \\
\textbf{Qwen-3-14B-GRPO (ours)} & \textbf{0.73} & 0.71 \\
\midrule
\end{tabular}
\end{table}

\subsection{Test Set Evaluation}\label{subsec:results_dataset_evaluation}

\subsubsection{\textbf{Strategy Model:}}

Table~\ref{tab:model_strategy_explanation} shows the results of Pen-Strategist evaluated on the test set using the setup described in Section~\ref{subsec:exp_setup_A}. As observed, predicting pentesting strategies remains challenging for current large language models, as reflected in consistently low GEval scores, mostly below 0.6. The strongest baseline, Claude-4.5-Sonnet, achieves 0.65 for strategy and 0.72 for explanation, yet still falls short of the proposed approach. It is also notable that more recent commercial model versions generally outperform their earlier counterparts (e.g., Claude 4.5 Sonnet vs. Claude 3 Haiku), suggesting incremental improvements with model evolution.

GRPO fine-tuning significantly enhances the performance of Qwen-3-14B on both strategy and explanation similarity, with scores increasing from 0.39 to 0.73 and from 0.45 to 0.71, respectively. The fine-tuned model outperforms all evaluated commercial models in strategy prediction, achieving an improvement of approximately 87\% over its base version. Notably, it surpasses Claude 4.5 Sonnet, the strongest commercial baseline, by 8 percentage points in strategy similarity, despite being considerably smaller in scale. In Figure~\ref{fig:strategy_example}, we present an example illustrating how GRPO training has improved the base model. The base model provides a high-level answer, whereas the fine-tuned model offers a more specific and actionable strategy by leveraging the SQL injection to achieve direct code execution through a webshell.

The explanation similarity of the Strategy model is on par with Claude-4.5-sonnet. Since the reasoning involves a step-by-step process with dense security-related terminology, an area where Claude models are typically stronger, the GEval score is slightly higher than that of the Strategy model.


\begin{table}[t]
\footnotesize
\centering
\caption{Step and MCP server prediction performance.}
\label{tab:step_mcp_results}
\begin{tabular}{lcccc}
\hline
\textbf{Model} & \textbf{Step} & \textbf{Step} & \textbf{MCP} & \textbf{MCP} \\
 & \textbf{Acc. (\%)} & \textbf{Micro F1} & \textbf{Acc. (\%)} & \textbf{Micro F1} \\
\hline
Claude 3 Haiku & 55.43 & 0.53 & 16.13 & 0.39 \\
Claude 4.5 sonnet & 63.44 & 0.66 & 24.73 & 0.60 \\
Gemini 2.0 Flash & 61.29 & 0.67 & 24.17 & 0.59 \\
Gemini 2.5 Flash & 64.52 & 0.68 & 30.11 & 0.62 \\
GPT-3.5-turbo & 41.76 & 0.45 & 18.52 & 0.40 \\
GPT 4.1 & 62.78 & 0.65 & 28.41 & 0.58 \\
GPT-4o-mini & 54.83 & 0.54 & 28.69 & 0.56 \\
GPT-5-mini & 59.66 & 0.65 & 33.81 & 0.50 \\
LLaMA-3.1-8B & 51.09 & 0.54 & 13.98 & 0.38 \\
\textbf{Step Model (ours)} & \textbf{82.87} & \textbf{0.80} & \textbf{48.88} & \textbf{0.64} \\
\hline
\end{tabular}
\end{table}

\begin{figure}[h]
\centering
\footnotesize

\begin{tcolorbox}[
    colback=lightbluebox,
    colframe=black,
    width=\linewidth
]

\textbf{Scenario:} \\
Reconnaissance identified key findings supporting SQL injection exploitation: an Apache web server hosting the “Stark Hotel” application and a dynamic parameter (\texttt{cod}) in the booking feature, which was confirmed to be vulnerable to union-based SQL injection. Exploitation of this vulnerability already demonstrates significant database interaction capabilities, including retrieval of the database version and user (\texttt{DBadmin@localhost}). More importantly, the ability to read system files confirms that the database has elevated privileges and supports file operations. Further inspection of the Apache configuration reveals the DocumentRoot (\texttt{/var/www/html}), and the environment indicates that MySQL file-writing functionality (e.g., \texttt{INTO OUTFILE}) is likely available.\\

\textbf{Base Model:} \\
Use the confirmed SQL injection to extract more database information, look for user credentials, exploit file writing capability to gain persistent access or execute commands, check for other vulnerabilities, and consider privilege escalation.\\

\textbf{Fine-tuned Model:} \\
The next strategy involves exploiting the confirmed SQL injection vulnerability to write a PHP webshell into the server's DocumentRoot using the \texttt{INTO OUTFILE} clause. This will allow us to execute arbitrary commands on the server. The SQL injection payload will be crafted to write a reverse shell or webshell into \texttt{/var/www/html}, which is the DocumentRoot. Once the shell is uploaded, it can be accessed via the web browser, granting a shell session on the target system.

\end{tcolorbox}

\caption{A sample strategy derived by Qwen-3-14B base model and the fine-tuned model (i.e., Pen-Strategist-Strategy model) for a pentesting scenario.}
\label{fig:strategy_example}

\end{figure}

\begin{figure}[b]
    \centering
    \includegraphics[width=\linewidth]{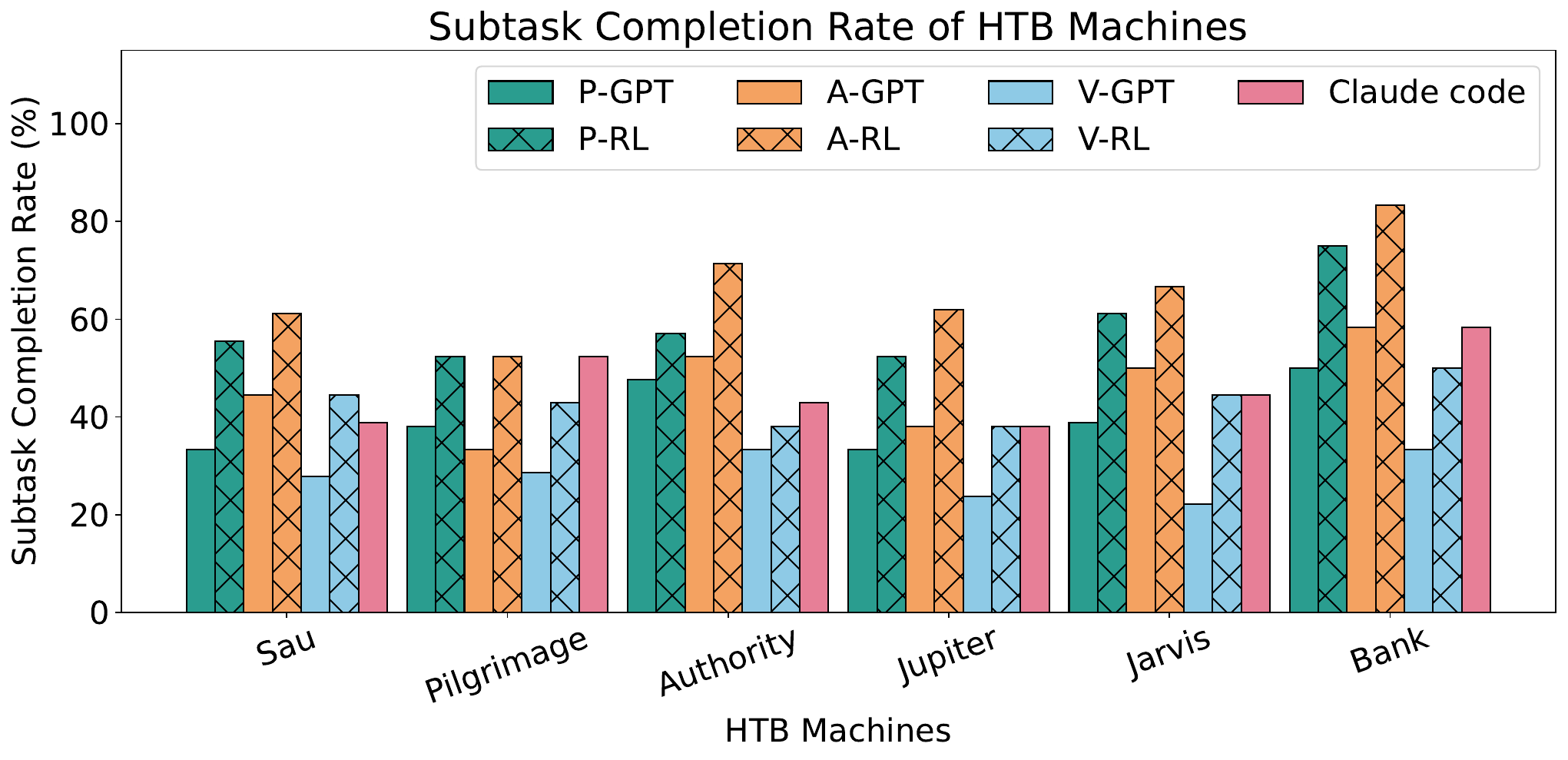}
    \caption{Subtask completion rates for HTB machines. P, A, and V denote PentestGPT, AutoPentester, and VulnBot, respectively. X-GPT uses the GPT-5 as the backend for the strategy analyzer, and X-RL uses the Strategy model.}
    \label{fig:htb_results}
\end{figure}

\begin{table*}[t]
\footnotesize
\centering
\caption{Performance of different LLMs on CTF-Known benchmark. Note that finetuning has significantly improved the Qwen model's performance.}
\label{tab:ctf_known}
\begin{threeparttable}
\begin{tabular}{lccccccc}
\toprule
\textbf{Model/Human} & \textbf{Web (218)} & \textbf{Pwn (459)} & \textbf{Misc (332)} & \textbf{Crypto (638)} & \textbf{Reverse (128)} & \textbf{Forensics (221)} & \textbf{Total (1996)} \\
\midrule

GPT-3.5-Turbo        & 72.02 & 79.08 & 82.23 & 81.03 & 79.69 & 76.47 & 79.21 \\
GPT-4-Turbo          & 80.73 & 85.40 & 87.95 & 86.21 & 90.62 & 85.07 & 85.87 \\
GPT-4o               & 83.03 & 88.02 & 89.46 & 88.56 & 89.84 & 86.43 & 87.83 \\
GPT-5-mini           & 88.26 & 91.21 & 92.41 & 93.88 & 93.43 & 91.67 & 92.26 \\
Llama-3-70B          & 83.03 & 86.06 & 88.86 & 85.89 & 89.84 & 87.33 & 86.52 \\
Gemini-2.5-Flash     & 79.58 & 79.20 & 86.22 & 81.50 & 82.41 & 78.49 & 81.23 \\
Claude-4.5-Sonnet    & 85.71 & 75.40 & 89.19 & 69.64 & 76.19 & 72.22 & 76.10 \\
\hline
Qwen-3-14B           & 67.69 & 76.03 & 74.87 & 78.60 & 72.59 & 64.37 & 72.38 \\
Qwen-3-14B-GRPO (Our)     & 81.31 & 85.19 & 81.96 & 87.37 & 89.47 & 89.29 & 85.71 \\

\bottomrule
\end{tabular}
\end{threeparttable}
\end{table*}

\subsubsection{\textbf{Step Model:}}

Next, we evaluate the trained Step model on the test dataset. As shown in Table~\ref{tab:step_mcp_results}, the Step model achieves the best performance across all evaluation metrics, clearly outperforming all baseline LLMs for both step and MCP prediction. Among the baselines, Gemini 2.5 Flash emerges as the strongest overall competitor, although it still lags behind the trained model by a noticeable margin (18\% in Step F1-score and 5\% in MCP F1-score). This consistent gap indicates that, while general-purpose LLMs can provide reasonable performance, they are not yet reliable for precise step and MCP server prediction in this setting. Overall, the results demonstrate that task-specific supervised training with a dedicated classifier is substantially more effective than prompting general-purpose LLMs, leading to more reliable structured predictions for automated execution.

\begin{table}[b]
\footnotesize
\centering
\caption{PicoCTF challenge results using different LLMs as the strategy analyzer in PentestGPT. We conduct 5 runs for each challenge and report the number of successful completions. The Qwen model used is Qwen-3-14B.}
\label{tab:picoctf_results}
\begin{tabular}{llccc}
\toprule
\textbf{Challenge} & \textbf{Category} & \makecell{\textbf{GPT-5}} & \makecell{\textbf{Qwen}\\\textbf{Base}} & \makecell{\textbf{Qwen-}\\\textbf{RL}} \\
\midrule
login                  & web             & 3  & 1 & 3 \\
advance-potion-making  & forensics       & 2  & 1 & 2 \\
spelling-quiz          & crypto          &  1 & 1 & 2 \\
caas                   & web             & 2  & 0 & 1 \\
XtrOrdinary            & crypto          & 2  & 1 & 1 \\
tripplesecure          & crypto          & 1  & 1 & 2 \\
clutteroverflow        & binary          &  1 & 0 & 1 \\
not                    & crypto  & 0  & 0 & 1 \\
scrambled-bytes        & forensics       & 0  & 0 & 0 \\
breadth                & reverse         & 0  & 0 & 0 \\
notepad                & web             & 0  & 0 & 0 \\
college-rowing-team    & crypto          &  1 & 0 & 2 \\
fermat-strings         & binary          &  0 & 0 & 1 \\
corrupt-key-1          & crypto          &  0 & 0 & 1 \\
SaaS                   & binary          &  0 & 0 & 0 \\
riscy business         & reverse         & 0  & 0 & 0 \\
homework               & binary          &  0 & 0 & 0 \\
lockdown-horses        & binary          &  0 & 0 & 0 \\
corrupt-key-2          & crypto          &  0 & 0 & 0 \\
vr-school              & binary          &  0 & 0 & 0 \\
\hline

Total & & 13 & 5 & 17 \\
\bottomrule
\end{tabular}
\end{table}

\subsection{Evaluate on Pentesting Frameworks}\label{subsec:results_setup_B}


As outlined in Section~\ref{subsec:setup B}, we assess the Strategy model by integrating it as the backend of strategy analyzer agents within automated pentesting frameworks. Figure~\ref{fig:htb_results} shows the performance of these frameworks across six HTB machines, using GPT-5 and the Strategy model as backends. The higher patterned bars show that our model consistently outperforms GPT-5, achieving higher subtask completion rates across all machines and frameworks. Overall, the finetuned model improved the subtask completion rate of PentestGPT by 46.5\% and AutoPentester by 43.4\%, VulnBot by 52.5\%, across all the test machines. These findings indicate that RL-augmented frameworks demonstrate clear performance gains, indicating that the fine-tuned model significantly enhances subtask completion by enabling more accurate next strategy derivation.

For completeness, we repeat the experiment using the Claude Code as the agentic pentester. We observed that Claude Code falls short compared to AutoPentester and PentestGPT with the Strategy model backend, achieving 45.83\% on average across machines. However, it is worth noting that Claude Code performs marginally better than VulnBot version with the fine-tuned model. 

\subsection{Cross-Task Generalization }\label{subsec:results_setup_c}

As outlined in Section~\ref{subsec:setup_c}, we assess the performance of the Strategy model on CTF challenges. The results in Table~\ref{tab:ctf_known} on the CTFKnown benchmark show that GRPO finetuning substantially enhances the Strategy model's reasoning and problem-solving performance across all task categories. The Qwen-3-14B base model achieves a total success rate of 72.38\%, which is substantially lower than GPT-4o and GPT-4-Turbo. This performance gap is expected, as commercial models benefit from significantly larger parameter scales and are trained on extensive proprietary datasets comparable to those used in the experiment. However, after GRPO fine-tuning, Qwen-3-14B-GRPO improves to 85.71\%, narrowing the gap with GPT-4o to just 2.12 percentage points and outperforming several strong baselines such as Llama-3-70B. Importantly, this gain was achieved with a much smaller 14B parameter model, and it can be deployed and hosted locally, offering a practical advantage in terms of cost and data privacy. Overall, fine-tuning consistently boosts performance across all domains, with the largest improvements observed in reasoning-heavy tasks such as reverse engineering and forensics.  


 Furthermore, Table~\ref{tab:picoctf_results} presents the results of the PicoCTF challenges conducted using the PentestGPT framework and different LLMs as the strategy analyzer. The results show a clear improvement in the Qwen model after fine-tuning. The base Qwen-3-14B achieves only 5 total successful cases, whereas the fine-tuned Qwen-RL reaches 17. Notably, Qwen-RL slightly outperforms GPT-5 in total successful attempts, indicating that fine-tuning improves the model’s ability to generalize reasoning across CTF challenges. 

\subsection{Ablation Study}\label{subsec:results_setup_D}


Our ablation study consists of evaluating the impact of the two newly introduced models, namely the strategy and step models, on pentesting tasks, with the results reported in Table~\ref{tab:ablation_subtask_completion}. In addition, we analyze the failure cases observed in the ablation study and present them in Table~\ref{tab:error_analysis}. The AutoPentester with the Strategy model (RL-A-Strategy) consistently improves subtask completion over the baseline AutoPentester (A) across all the machines, gaining 21.8\%.  This is further supported by Table~\ref{tab:error_analysis}, where “incorrect strategy selected” errors are reduced from 4 to 2, indicating that the Strategy model improves reasoning at the planning level.
With the addition of the step prediction module (RL-A-Strategy + Step), performance further increases by 11.5\% on average, due to stable strategy execution. Table~\ref{tab:error_analysis} shows that tool-related errors are fully eliminated, indicating that the step model effectively guides correct tool selection. Overall, the Strategy model improves correct strategy selection, while the step model reduces tool-selection errors and strengthens execution reliability, leading to the best overall performance as measured by subtask completion rate. 

\begin{table}[t]
\footnotesize
\centering
\caption{Ablation study based on subtask completion (\%) on HTB machines to measure the impact of fine-tuned models.}
\label{tab:ablation_subtask_completion}
\begin{tabular}{llc}
\toprule
\textbf{Machine} & \textbf{Method} & \textbf{Subtask Completion (\%)} \\
\midrule
\multirow{3}{*}{Sau (6 subtasks)} 
& A                    & 44.44 \\
& RL-A-Strategy        & 55.58 \\
& RL-A-Strategy + Step & \textbf{61.11} \\
\midrule
\multirow{3}{*}{Authority (7 subtasks)} 
& A                    & 52.38 \\
& RL-A-Strategy        & 61.90 \\
& RL-A-Strategy + Step & \textbf{71.42} \\
\midrule
\multirow{3}{*}{Jarvis (6 subtasks)} 
& A                    & 50.00 \\
& RL-A-Strategy        & 61.11 \\
& RL-A-Strategy + Step & \textbf{66.67} \\
\bottomrule
\end{tabular}
\end{table}

\begin{table}[t]
\footnotesize
\centering
\caption{Error analysis in the ablation study.}
\label{tab:error_analysis}
\begin{tabular}{lccc}
\toprule
\textbf{Error Type} & \textbf{A} & \textbf{RL-A-Strategy} & \textbf{RL-A-Str + Step} \\
\midrule
Try tools which are not installed & 1 & 1 & 0 \\
Try GUI-based tools (not supported)               & 1 & 2 & 0 \\
Incorrect strategy selected       & 4 & 2 & 2 \\
Cannot craft the valid exploit    & 2 & 1 & 3 \\
Other                             & 1 & 3 & 4 \\
\bottomrule
\end{tabular}
\end{table}

\subsection{Survey Analysis}\label{subsec:results_setup_E}

\begin{figure}[t]
    \centering
    \includegraphics[width=\linewidth]{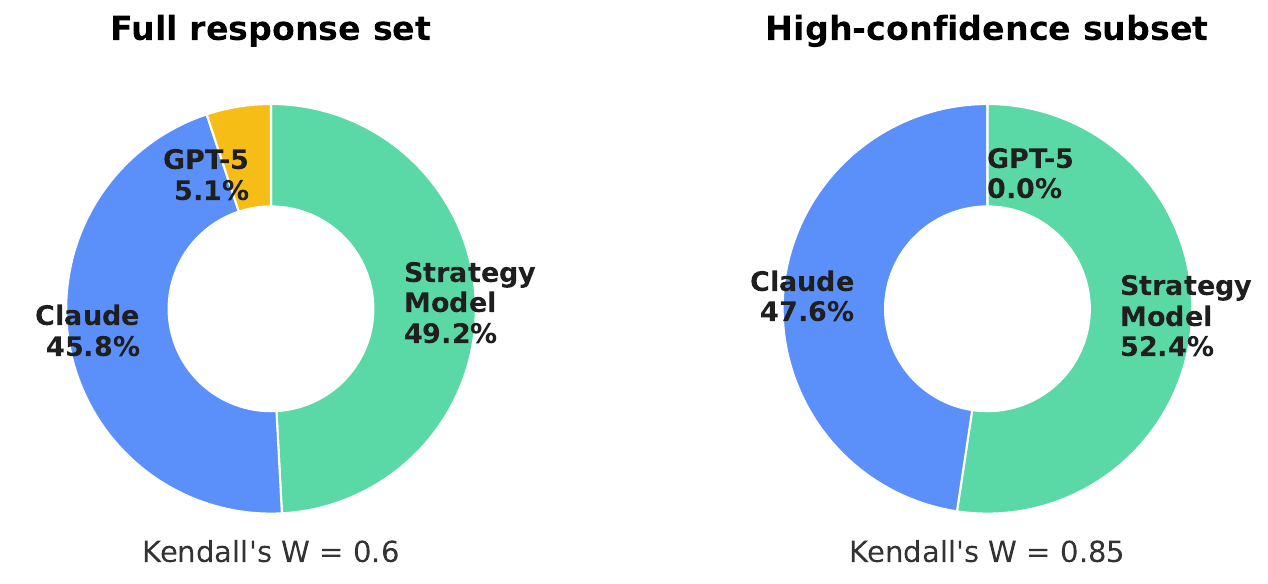}
    \caption{Survey results analysis.}
    \label{fig:survey_results}
\end{figure}

As described in Section~\ref{subsec:setup_e}, we conducted a user study with 12 cybersecurity experts to qualitatively evaluate the Pen-Strategist. As illustrated in Figure~\ref{fig:survey_results}, across the full response set, the Strategy model attains a higher first-choice rate than Claude-4.6-Sonnet, exceeding it by 3.4 percentage points, and inter-participant agreement is moderately strong, with a Kendall’s W of 0.6. When focusing on the high-confidence subset, the Strategy model’s advantage becomes more pronounced, achieving a 4.8 percentage point lead in first-choice rate. In this subset, agreement also rises to 0.85, suggesting a clear and confident preference for the Strategy model over commercial LLMs. Notably, GPT-5 performs significantly worse under both settings. Overall, the findings consistently show that the Strategy model is the most preferred by experts, with Claude performing competitively.



Below we show some example free text comments we received from the participants. Here, Options 1, 2, and 3 refer to Strategy, Claude-4.6-sonnet, and GPT-5 models, respectively. 

\noindent{\textbf{Comment 1:} \textit{"Option 2 includes several redundant checks that can be avoided, such as SUID/SGID bit checking, since we have already identified the execution behavior and privilege context of test.py and test.txt. Option 3 lacks confidence in its explanation and is less precise. It also suggests gaining script manager privileges, even though we already have access to that user. Option 1 correctly understands the key observation that test.py can be modified and is likely executed with root privileges (e.g., via a cron job). Therefore, it provides a clear and effective path to escalate privileges by modifying the script."}} \\ \vspace{-2mm}

\noindent{\textbf{Comment 2:} \textit{Option 2 identifies a strong strategy, however focuses predominantly on the enumeration of the API. Option 1 expands upon Option 2, and provides context about the importance of researching CVEs relevant to the current stack.}} \\ \vspace{-2mm}

Both the comments favor our model for its clarity and stronger strategic reasoning. In Comment 1, our model is credited with correctly identifying the key exploitation path, modifying a writable script likely executed with elevated privileges, while Claude-4.6-sonnet is criticized for redundant checks and GPT-5 for less confident and imprecise reasoning. In Comment 2, Claude-4.6-sonnet demonstrates a valid approach; however, it is viewed as overly focused on enumeration, whereas our model extends this by incorporating broader contextual awareness, such as the importance of CVE research. Overall, participants emphasize our model’s ability to combine accurate technical insight with actionable, context-aware strategy compared to the other two models.

\begin{figure}[b]
    \centering
    \includegraphics[width=\linewidth]{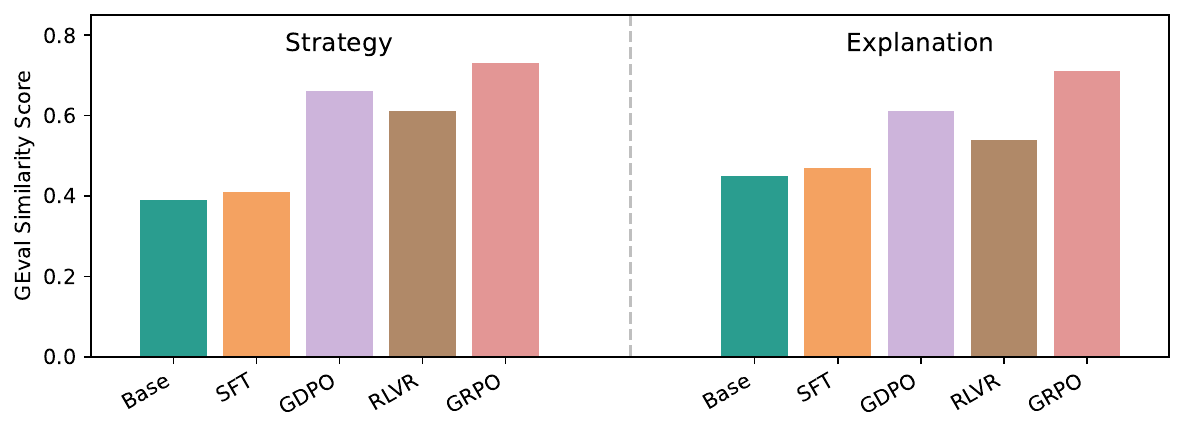}
    \caption{Evaluation of different training approaches for strategy generation compared to Qwen-3-14B base model, measured using GEval similarity scores.}
    \label{fig:training_approaches}
\end{figure}

\subsection{Extended Experiments}


To further evaluate the Pen-Strategist, we conduct extended experiments. 

\subsubsection{\textbf{Evaluating Different Fine-tuning Techniques:}} Here, we assess how different training strategies affect the quality of generated pentesting strategies and their explanations. As shown in Figure~\ref{fig:training_approaches}, reinforcement learning–based methods consistently outperform supervised fine-tuning (SFT) across both metrics. Among the reinforcement learning techniques, GRPO achieves the best performance, reaching 0.73 for strategy generation and 0.71 for explanation quality compared to GDPO and RLVR. SFT marginally improves the performance of the base model, indicating that naive supervised fine-tuning is insufficient for this task. This can be attributed to SFT shifting the model weights too far from their pretrained initialization, which may lead to overfitting to the training set, reduced output diversity, and diminished generalization ability for the test set~\cite{bachmann2024pitfalls, holtzmancurious}. Overall, these results demonstrate the effectiveness of reinforcement learning approaches, with GRPO in particular providing the most substantial improvements in both strategic reasoning and explanation quality.

\begin{table}[t]
\footnotesize
\centering
\caption{Performance comparison of base and fine-tuned models for strategy generation.}
\label{tab:fine-tune_comparison}
\begin{tabular}{lcccc}
\toprule
\multirow{2}{*}{\textbf{Model}} & \multicolumn{2}{c}{\textbf{Base}} & \multicolumn{2}{c}{\textbf{Fine-tuned}} \\
\cmidrule(lr){2-3} \cmidrule(lr){4-5}
 & \textbf{Strategy} & \textbf{Expl.} & \textbf{Strategy} & \textbf{Expl.} \\
\midrule
Qwen-3-8B-GRPO & 0.16 & 0.32 & 0.33 & 0.50 \\
Nemotron-cascade-14B & 0.45 & 0.45 & 0.64 & 0.56 \\
Mistral-3-14B & 0.47 & 0.45 & 0.63 & 0.75 \\
Qwen-3-14B-GRPO & 0.39 & 0.45 & 0.74 & 0.72 \\
\bottomrule
\end{tabular}
\end{table}

\begin{table}[t]
\footnotesize
\centering
\caption{Performance comparison across models using Pass@k metrics. Here, Str. and Exp. represents GEval scores for strategy and the explanation.}
\label{tab:passk_results}
\begin{tabular}{lcccccc}
\toprule
\multirow{2}{*}{\textbf{Model}} & \multicolumn{2}{c}{\textbf{Pass@1}} & \multicolumn{2}{c}{\textbf{Pass@3}} & \multicolumn{2}{c}{\textbf{Pass@5}} \\
\cmidrule(lr){2-3} \cmidrule(lr){4-5} \cmidrule(lr){6-7}
 & \textbf{Str.} & \textbf{Exp.} & \textbf{Str.} & \textbf{Exp.} & \textbf{Str.} & \textbf{Exp.} \\
\midrule
Claude 4.5 Sonnet & 0.65 & 0.72 & 0.73 & 0.74 & 0.75 & 0.74 \\
Gemini 2.5 Flash & 0.45 & 0.56 & 0.50 & 0.60 & 0.55 & 0.63 \\
GPT-5 & 0.62 & 0.63 & 0.72 & 0.70 & 0.73 & 0.72 \\
Qwen-3-14B-GRPO (ours) & \textbf{0.73} & \textbf{0.71} & \textbf{0.75} & \textbf{0.74} & \textbf{0.77} & \textbf{0.76} \\
\bottomrule
\end{tabular}
\end{table}

\subsubsection{\textbf{Fine-tuning Different Open-source Models:}}

To generalize the effectiveness of the GRPO training using the collected dataset, we compare the performance of fine-tuned Qwen-3-8B~\cite{Qwen3_8B}, Nemotron-cascade-14B~\cite{nemotron_14B}, and Mistral-3-14B-reasoning~\cite{ministral-3-14b-reasoning} models with their respective base models (similar to the experiment setup in Section~\ref{subsec:exp_setup_A}). As presented in Table~\ref{tab:fine-tune_comparison}, all models benefit from fine-tuning, with Qwen-3-14B demonstrating superior overall performance across both strategy generation and explanation quality.
Other models also show consistent gains, averaging 60.7\% in strategy generation and 49.6\% in explanation quality. Overall, GRPO fine-tuning substantially improves performance across architectures.

\subsubsection{\textbf{Pass@k Evaluation:}} Finally, we evaluate the strategy generation performance of different models on the test set of the dataset using Pass@k metrics, where \(k\) represents the number of strategies generated for a sample. For instance, Pass@3 refers to generating 3 strategies and selecting the best one for GEval calculations. Table~\ref{tab:passk_results} shows that our Strategy consistently outperforms strong commercial baselines across all \(k\) values. Overall, performance increases across all models with \(k\), as generating more candidate strategies increases the likelihood of including a correct solution. The superior performance of the Strategy model across all the k-cases demonstrates consistency in generating correct strategies and explanations compared to other models. Furthermore, unlike other models, the Strategy model achieves a higher Pass@1 rate, indicating its ability to generate the correct strategy in a single attempt.

\section{Discussion and Concluding Remarks}

We propose Pen-Strategist, a framework to derive strategies through logical reasoning for penetration testing scenarios and predict the actions and MCP servers to execute the selected strategy. To achieve that, we collect a reasoning dataset using HTB and Vulnhub machines and fine-tune an open-source Qwen-3-14B model for strategy derivation using GRPO. Furthermore, we train a semantic-based dual-head CNN classifier to predict the next step and the MCP servers. The extensive experiments conducted demonstrate that fine-tuning improves the strategy derivation performance of the open-source model by 87\%. Furthermore, the step model achieves 82.8\% accuracy in step prediction, outperforming commercial LLMs. When integrated into frameworks such as PentestGPT, the combined strategy and step models improve subtask completion rates on HTB machines, highlighting their practical effectiveness. We further demonstrate that fine-tuning also boosts performance in related red teaming tasks, including CTF challenges, indicating strong generalization. Finally, a user study reveals that security professionals prefer the strategies generated by our model over Claude-4.6-Sonnet by a margin of 4.8\%. Next, we discuss the general comments, limitations, and future work.

\noindent{\textbf{Local Models for Data Privacy: } The Pen-Strategist framework can be locally deployed and enables on-premise pentesting strategy and step formulation without exposing sensitive system information or vulnerabilities to third-party LLMs. This significantly enhances data privacy for enterprises, addressing a key concern in automated pentesting. More specifically, the Qwen-3-14B model we finetuned can be deployed on a single GPU with 80GB VRAM, representing a substantially lower hardware requirement compared to larger models. However, results from HTB experiments indicate that even fine-tuned LLMs continue to struggle with identifying correct strategies in certain pentesting tasks, primarily due to limited model scale and insufficient training on relevant data and scenarios. Scaling to larger open-source models, such as Qwen-3-235B~\cite{alibaba_cloud_2025a}, and fine-tuning them on more comprehensive datasets is likely to yield improved performance. In such cases, however, local deployment will require more compute.}

\noindent{\textbf{MCP Classifier and Recent Developments:}} In our approach, the step model functions as both an action predictor and an MCP server classifier, helping to minimize incorrect tool usage. Alternatively, one could define Claude Code skills~\cite{agent_skills_2025} that guide an LLM to execute the chosen steps directly, removing the need for a separate MCP selection classifier. However, this shifts full responsibility to the LLM’s decision-making, which may introduce errors, as reflected by the relatively low F1 scores of other LLMs in Table~\ref{tab:step_mcp_results}. Furthermore, it may hallucinate the use of unavailable tools or may try to install them, and therefore requires guardrails to ensure safe execution and protect privacy within the environment. A similar behavior was observed in the recent OpenClaw~\cite{openclaw_2025} agent, where the system incurs substantial token costs due to looping in incorrect strategies and attempting to use unavailable tools. In contrast, our approach provides a more constrained and structured environment covering the most commonly used pentesting tools, while still allowing flexibility to add more based on the specific needs of the pentester.

\noindent{\textbf{Dataset Extension:}} Our dataset currently includes samples from only 240 vulnerable machines, primarily due to constraints in time and human effort. However, by releasing it publicly, we enable others to expand it with additional human-curated entries following the same format, thereby improving its scale and utility for training further reasoning models for security. Moreover, this approach can be extended to develop more general red and blue teaming models by incorporating reasoning data from related tasks such as log analysis, digital forensics, and software security analysis.

In conclusion, although much of the existing work emphasizes agentic systems for automating security tasks, a major limitation is the model’s ability to reason logically and systematically about current findings to determine the next strategy. Another challenge lies in executing these strategies while respecting the constraints of the execution environment. To this end, we proposed Pen-Strategist, consisting of two models, strategy and step, which together lead to significant performance improvements in pretesting strategy formation and across a range of security tasks.




\bibliographystyle{ACM-Reference-Format}
\bibliography{Paper/Sections/bibliography}

@inproceedings{deng2024pentestgpt,
  title        = {{PentestGPT}: Evaluating and harnessing large language models for automated penetration testing},
  author       = {Deng, Gelei and Liu, Yi and Mayoral-Vilches, V{\'\i}ctor and Liu, Peng and Li, Yuekang and Xu, Yuan and Zhang, Tianwei and Liu, Yang and Pinzger, Martin and Rass, Stefan},
  booktitle    = {33rd USENIX Security Symposium (USENIX Security 24)},
  pages        = {847--864},
  year         = {2024}
}

@inproceedings{ginige2025autopentester,
  title        = {Autopentester: An {LLM} agent-based framework for automated pentesting},
  author       = {Ginige, Yasod and Niroshan, Akila and Jain, Sajal and Seneviratne, Suranga},
  booktitle    = {2025 IEEE 24th International Conference on Trust, Security and Privacy in Computing and Communications (TrustCom)},
  pages        = {163--174},
  year         = {2025},
  organization = {IEEE}
}

@inproceedings{shen2025pentestagent,
  title        = {Pentestagent: Incorporating {LLM} agents to automated penetration testing},
  author       = {Shen, Xiangmin and Wang, Lingzhi and Li, Zhenyuan and Chen, Yan and Zhao, Wencheng and Sun, Dawei and Wang, Jiashui and Ruan, Wei},
  booktitle    = {Proceedings of the 20th ACM Asia Conference on Computer and Communications Security},
  pages        = {375--391},
  year         = {2025}
}

@article{kong2025vulnbot,
  title        = {Vulnbot: Autonomous penetration testing for a multi-agent collaborative framework},
  author       = {Kong, He and Hu, Die and Ge, Jingguo and Li, Liangxiong and Li, Tong and Wu, Bingzhen},
  journal      = {arXiv preprint arXiv:2501.13411},
  year         = {2025}
}

@inproceedings{CTFKnow,
  title        = {Measuring and augmenting large language models for solving capture-the-flag challenges},
  author       = {Ji, Zimo and Wu, Daoyuan and Jiang, Wenyuan and Ma, Pingchuan and Li, Zongjie and Wang, Shuai},
  booktitle    = {Proceedings of the 2025 ACM SIGSAC Conference on Computer and Communications Security},
  pages        = {603--617},
  year         = {2025}
}

@misc{anthropic_2026,
  author       = {Anthropic},
  title        = {Claude Code},
  howpublished = {\url{https://code.claude.com/docs/en/quickstart}},
  year         = {2026}
}

@article{AutoAttacker,
  title        = {{AutoAttacker:} A large language model guided system to implement automatic cyber-attacks},
  author       = {Xu, Jiacen and Stokes, Jack W and McDonald, Geoff and Bai, Xuesong and Marshall, David and Wang, Siyue and Swaminathan, Adith and Li, Zhou},
  journal      = {arXiv preprint arXiv:2403.01038},
  year         = {2024}
}

@inproceedings{hu2020automated,
  title        = {Automated penetration testing using deep reinforcement learning},
  author       = {Hu, Zhenguo and Beuran, Razvan and Tan, Yasuo},
  booktitle    = {IEEE European Symposium on Security and Privacy Workshops (EuroS\&PW)},
  pages        = {2--10},
  year         = {2020}
}

@article{team2023gemini,
  title        = {Gemini: A family of highly capable multimodal models},
  author       = {Team, Gemini and Anil, Rohan and Borgeaud, Sebastian and Alayrac, Jean-Baptiste and others},
  journal      = {arXiv preprint arXiv:2312.11805},
  year         = {2023}
}

@article{brown2020language,
  title        = {Language models are few-shot learners},
  author       = {Brown, Tom and Mann, Benjamin and Ryder, Nick and Subbiah, Melanie and Kaplan, Jared D and Dhariwal, Prafulla and Neelakantan, Arvind and Shyam, Pranav and others},
  journal      = {Advances in Neural Information Processing Systems},
  volume       = {33},
  pages        = {1877--1901},
  year         = {2020}
}

@misc{hackthebox_2024,
  author       = {{Hack The Box}},
  title        = {Hack The Box},
  howpublished = {\url{https://www.hackthebox.com/}},
  year         = {2024}
}

@article{wu2025SecRL,
  title        = {{ExCyTIn-Bench}: Evaluating {LLM} agents on cyber threat investigation},
  author       = {Wu, Yiran and Velazco, Mauricio and Zhao, Andrew and Luj{\'a}n, Manuel Ra{\'u}l Mel{\'e}ndez and Movva, Srisuma and Roy, Yogesh K and Nguyen, Quang and Rodriguez, Roberto and Wu, Qingyun and Albada, Michael and others},
  journal      = {arXiv preprint arXiv:2507.14201},
  year         = {2025}
}

@inproceedings{bhusal2024secure,
  title        = {{SECURE}: Benchmarking large language models for cybersecurity},
  author       = {Bhusal, Dipkamal and Alam, Md Tanvirul and Nguyen, Le and Mahara, Ashim and Lightcap, Zachary and Frazier, Rodney and Fieblinger, Romy and Torales, Grace Long and Blakely, Benjamin A and Rastogi, Nidhi},
  booktitle    = {2024 Annual Computer Security Applications Conference (ACSAC)},
  pages        = {15--30},
  year         = {2024},
  organization = {IEEE}
}

@article{shao2024deepseekmath,
  title        = {{DeepSeekMath}: Pushing the limits of mathematical reasoning in open language models},
  author       = {Shao, Zhihong and Wang, Peiyi and Zhu, Qihao and Xu, Runxin and Song, Junxiao and Bi, Xiao and Zhang, Haowei and Zhang, Mingchuan and Li, YK and Wu, Yang and others},
  journal      = {arXiv preprint arXiv:2402.03300},
  year         = {2024}
}

@article{liu2026gdpo,
  title        = {{GDPO}: Group reward-decoupled normalization policy optimization for multi-reward {RL} optimization},
  author       = {Liu, Shih-Yang and Dong, Xin and Lu, Ximing and Diao, Shizhe and Belcak, Peter and Liu, Mingjie and Chen, Min-Hung and Yin, Hongxu and Wang, Yu-Chiang Frank and Cheng, Kwang-Ting and others},
  journal      = {arXiv preprint arXiv:2601.05242},
  year         = {2026}
}

@article{wen2025rlvr,
  title        = {Reinforcement learning with verifiable rewards implicitly incentivizes correct reasoning in base {LLMs}},
  author       = {Wen, Xumeng and Liu, Zihan and Zheng, Shun and Ye, Shengyu and Wu, Zhirong and Wang, Yang and Xu, Zhijian and Liang, Xiao and Li, Junjie and Miao, Ziming and others},
  journal      = {arXiv preprint arXiv:2506.14245},
  year         = {2025}
}

@misc{ministral-3-14b-reasoning,
  author       = {{Mistral AI}},
  title        = {Ministral-3-14B-Reasoning-2512},
  howpublished = {\url{https://huggingface.co/mistralai/Ministral-3-14B-Reasoning-2512}},
  year         = {2025},
  month        = dec
}

@misc{Qwen3_8B,
  author       = {{Alibaba Cloud}},
  title        = {Qwen3-8B},
  howpublished = {\url{https://huggingface.co/Qwen/Qwen3-8B}},
  year         = {2025},
  month        = jun
}

@misc{Qwen3_14B,
  author       = {{Alibaba Cloud}},
  title        = {Qwen3-14B},
  howpublished = {\url{https://huggingface.co/Qwen/Qwen3-14B}},
  year         = {2025},
  month        = jun
}

@misc{nemotron_14B,
  author       = {{NVIDIA}},
  title        = {Nemotron-Cascade-14B-Thinking},
  howpublished = {\url{https://huggingface.co/nvidia/Nemotron-Cascade-14B-Thinking}},
  year         = {2026},
  month        = apr
}

@inproceedings{hu2022lora,
  title        = {{LoRA}: Low-rank adaptation of large language models},
  author       = {Hu, Edward J and Shen, Yelong and Wallis, Phillip and Allen-Zhu, Zeyuan and Li, Yuanzhi and Wang, Shean and Wang, Liang and Chen, Weizhu},
  booktitle    = {International Conference on Learning Representations},
  year         = {2022}
}

@inproceedings{geval,
  title        = {{G-Eval}: {NLG} evaluation using {GPT-4} with better human alignment},
  author       = {Liu, Yang and Iter, Dan and Xu, Yichong and Wang, Shuohang and Xu, Ruochen and Zhu, Chenguang},
  booktitle    = {Proceedings of the 2023 Conference on Empirical Methods in Natural Language Processing},
  pages        = {2511--2522},
  year         = {2023}
}

@article{kullback1951kullback,
  title        = {Kullback--{Leibler} divergence},
  author       = {Kullback, Solomon},
  journal      = {Encyclopedia of Machine Learning},
  pages        = {581--583},
  year         = {1951}
}

@article{lewis1971kendall,
  title        = {Kendall's Coefficient of Concordance for sociometric rankings with self excluded},
  author       = {Lewis, Gordon H and Johnson, Richard G},
  journal      = {Sociometry},
  pages        = {496--503},
  year         = {1971},
  publisher    = {JSTOR}
}

@article{bachmann2024pitfalls,
  title        = {The pitfalls of next-token prediction},
  author       = {Bachmann, Gregor and Nagarajan, Vaishnavh},
  journal      = {arXiv preprint arXiv:2403.06963},
  year         = {2024}
}

@inproceedings{holtzmancurious,
  title        = {The Curious Case of Neural Text Degeneration},
  author       = {Holtzman, Ari and Buys, Jan and Du, Li and Forbes, Maxwell and Choi, Yejin},
  booktitle    = {International Conference on Learning Representations},
  year         = {2020}
}

@misc{fedramp_2024,
  author       = {{FedRAMP}},
  title        = {{FedRAMP} Penetration Test Guidance},
  howpublished = {\url{https://www.fedramp.gov/assets/resources/documents/CSP_Penetration_Test_Guidance_public_comment.pdf}},
  year         = {2024},
  month        = apr
}

@inproceedings{al2018study,
  title        = {A study on penetration testing process and tools},
  author       = {Al Shebli, Hessa Mohammed Zaher and Beheshti, Babak D},
  booktitle    = {2018 IEEE Long Island Systems, Applications and Technology Conference},
  pages        = {1--7},
  year         = {2018}
}

@article{zhang2025cyberllama,
  title        = {{CyberLlama}: A fine-tuned large language model for cybersecurity named entity recognition},
  author       = {Zhang, Hao and Wu, Tingmin and Zhu, Tianqing and Wen, Sheng and Xiang, Yang},
  journal      = {Knowledge-Based Systems},
  pages        = {114183},
  year         = {2025},
  publisher    = {Elsevier}
}

@article{ginige2025trafficllm,
  title        = {{TrafficLLM}: {LLMs} for improved open-set encrypted traffic analysis},
  author       = {Ginige, Yasod and Silva, Bhanuka and Dahanayaka, Thilini and Seneviratne, Suranga},
  journal      = {Computer Networks},
  pages        = {111847},
  year         = {2025},
  publisher    = {Elsevier}
}

@article{dai2025qoq,
  title        = {{QoQ-Med}: Building multimodal clinical foundation models with domain-aware {GRPO} training},
  author       = {Dai, Wei and Chen, Peilin and Ekbote, Chanakya and Liang, Paul Pu},
  journal      = {arXiv preprint arXiv:2506.00711},
  year         = {2025}
}

@article{qiu2025open,
  title        = {{Open-Medical-R1}: How to choose data for {RLVR} training at medicine domain},
  author       = {Qiu, Zhongxi and Zhang, Zhang and Hu, Yan and Li, Heng and Liu, Jiang},
  journal      = {arXiv preprint arXiv:2504.13950},
  year         = {2025}
}

@article{liu2025fin,
  title        = {{Fin-R1}: A large language model for financial reasoning through reinforcement learning},
  author       = {Liu, Zhaowei and Guo, Xin and Yang, Zhi and Lou, Fangqi and Zeng, Lingfeng and Niu, Jinyi and Li, Mengping and Qi, Qi and Liu, Zhiqiang and Han, Yiyang and others},
  journal      = {arXiv preprint arXiv:2503.16252},
  year         = {2025}
}

@article{dai2025legal,
  title        = {Legal $\Delta$: Enhancing legal reasoning in {LLMs} via reinforcement learning with chain-of-thought guided information gain},
  author       = {Dai, Xin and Xu, Buqiang and Liu, Zhenghao and Yan, Yukun and Xie, Huiyuan and Yi, Xiaoyuan and Wang, Shuo and Yu, Ge},
  journal      = {arXiv preprint arXiv:2508.12281},
  year         = {2025}
}

@article{rafailov2023direct,
  title        = {Direct preference optimization: Your language model is secretly a reward model},
  author       = {Rafailov, Rafael and Sharma, Archit and Mitchell, Eric and Manning, Christopher D and Ermon, Stefano and Finn, Chelsea},
  journal      = {Advances in Neural Information Processing Systems},
  volume       = {36},
  pages        = {53728--53741},
  year         = {2023}
}

@misc{agent_skills_2025,
  author       = {Anthropic},
  title        = {Agent Skills Overview},
  howpublished = {\url{https://platform.claude.com/docs/en/agents-and-tools/agent-skills/overview}},
  year         = {2025}
}

@misc{openclaw_2025, title={GitHub - openclaw/openclaw: Your own personal AI assistant. Any OS. Any Platform. The lobster way.}, url={https://github.com/openclaw/openclaw}, journal={GitHub}, author={openclaw}, year={2025} }

@misc{alibaba_cloud_2025a, title={Qwen/Qwen3-235B-A22B · Hugging Face}, url={https://huggingface.co/Qwen/Qwen3-235B-A22B}, journal={Huggingface.co}, author={Alibaba Cloud}, year={2025}, month={Jun} }


\newpage

\appendix

\section{Open Science}
We published our dataset and codes in an anonymous GitHub repository (https://anonymous.4open.science/r/Pentest-Strategist-B783/). More specifically, it contains the following components.
\begin{itemize}
    \item The dataset
    \item Automated dataset collection code
    \item Training code for the strategy and the step models
    \item Codes for the experiments
    \item Anonymised survey results and analysis code.
\end{itemize}

The repository contains README files for each section, which guide users to set up the Python environment and run the experiments. Please follow those steps to execute the code successfully.

\section{Ethical Considerations}

The survey received formal approval from our University’s Ethics Committee following a comprehensive review process. Additionally, all responses were collected anonymously (without any personal details), and informed consent was obtained from participants via a dedicated form. As a result, the study fully complies with the University’s ethical guidelines, which are designed to international standards.

As this work presents a framework for strategy generation in penetration testing, supported by a dataset collected from simulated public platforms such as Hack-The-Box and VulnHub, we do not identify any potential risks to society arising from this research.






\section{Prompts}\label{app:prompts}

This section presents the complete prompts used during model fine-tuning and evaluation of the Pen-Strategist framework. 

\subsection*{System Prompt}

\begin{figure}[h]
    \centering
    \includegraphics[width=0.9\linewidth]{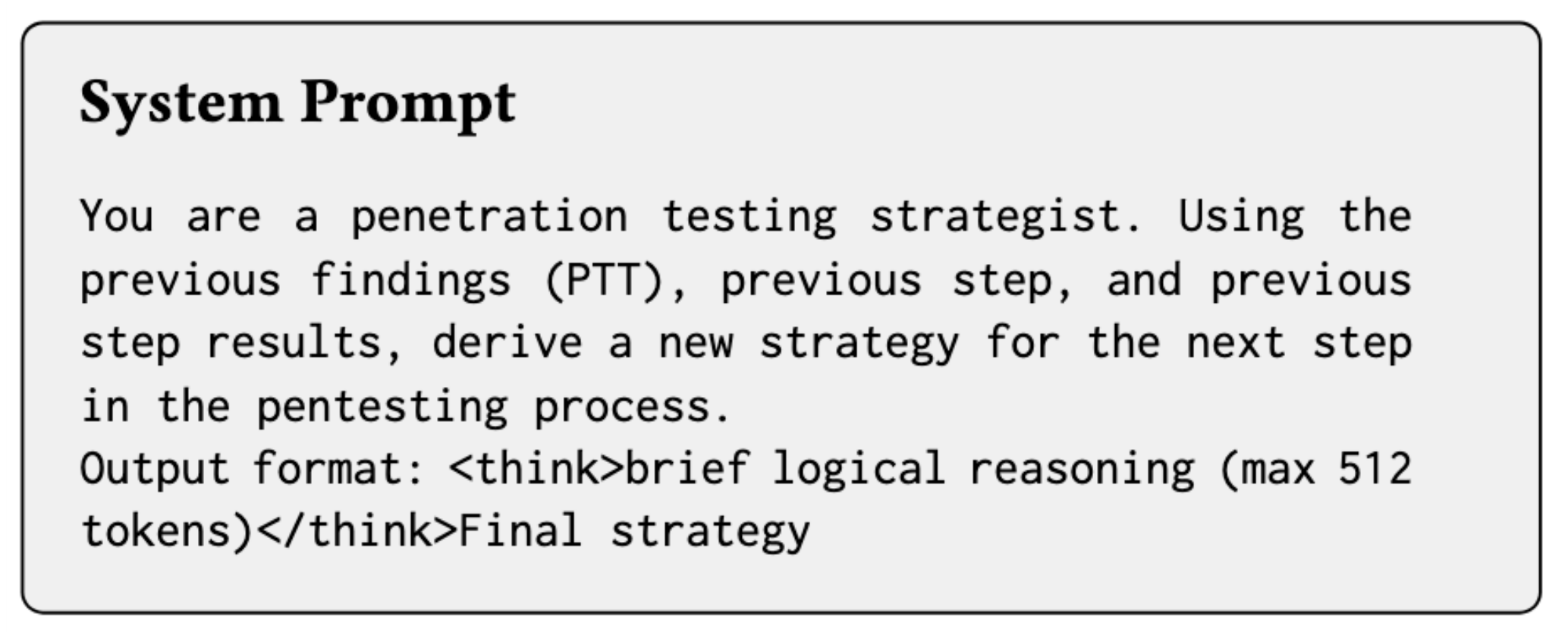}
    \caption{System prompt used for the Strategy model finetuning.}
    \label{fig:system_prompt}
\end{figure}

Figure~\ref{fig:system_prompt} presents the system prompt used during both fine-tuning and inference of the Strategy model. As described in Section~\ref{subsubsec:strategy_model}, the model is fine-tuned to act as a domain-specific penetration testing strategist. The system prompt instructs the model to derive a new strategy for the next pentesting step based on the current attack environment state, and specifies the required output format: a brief chain-of-thought reasoning enclosed in \texttt{<think>} tags followed by the final strategy. The reasoning phase is capped at 512 tokens, consistent with the generation length reward $R_l$ defined in Section~\ref{subsubsec:strategy_model}, which penalizes outputs exceeding the maximum token count to prevent reasoning explosion during training.

\subsection*{User Prompt}

\begin{figure}[h]
    \centering
    \includegraphics[width=0.9\linewidth]{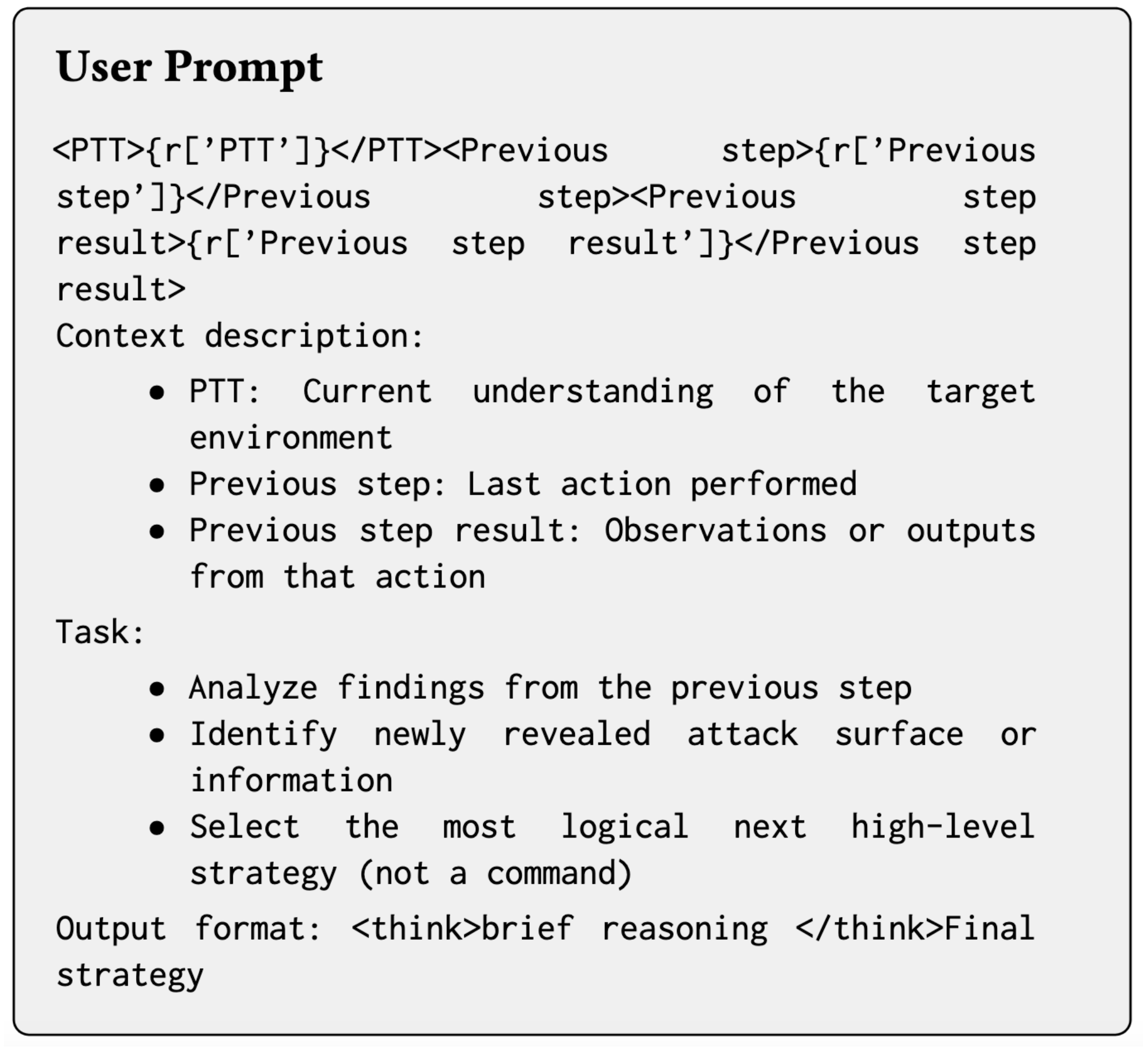}
    \caption{User prompt used for the Strategy model finetuning.}
    \label{fig:user_prompt}
\end{figure}

Figure~\ref{fig:user_prompt} presents the user prompt, which provides the structured input context for each training instance, as described in Section~\ref{subsubsec:dataset_fomalities}. Specifically, it supplies three input fields drawn from the dataset: the PenTest Tree (PTT), which summarizes the current attack state including previously executed steps and their findings; the Previous Step, representing the most recent action performed; and the Previous Step Result, containing the observed output of that action. 

\subsection*{Reward Model Prompt}

\begin{figure}[h]
    \centering
    \includegraphics[width=0.9\linewidth]{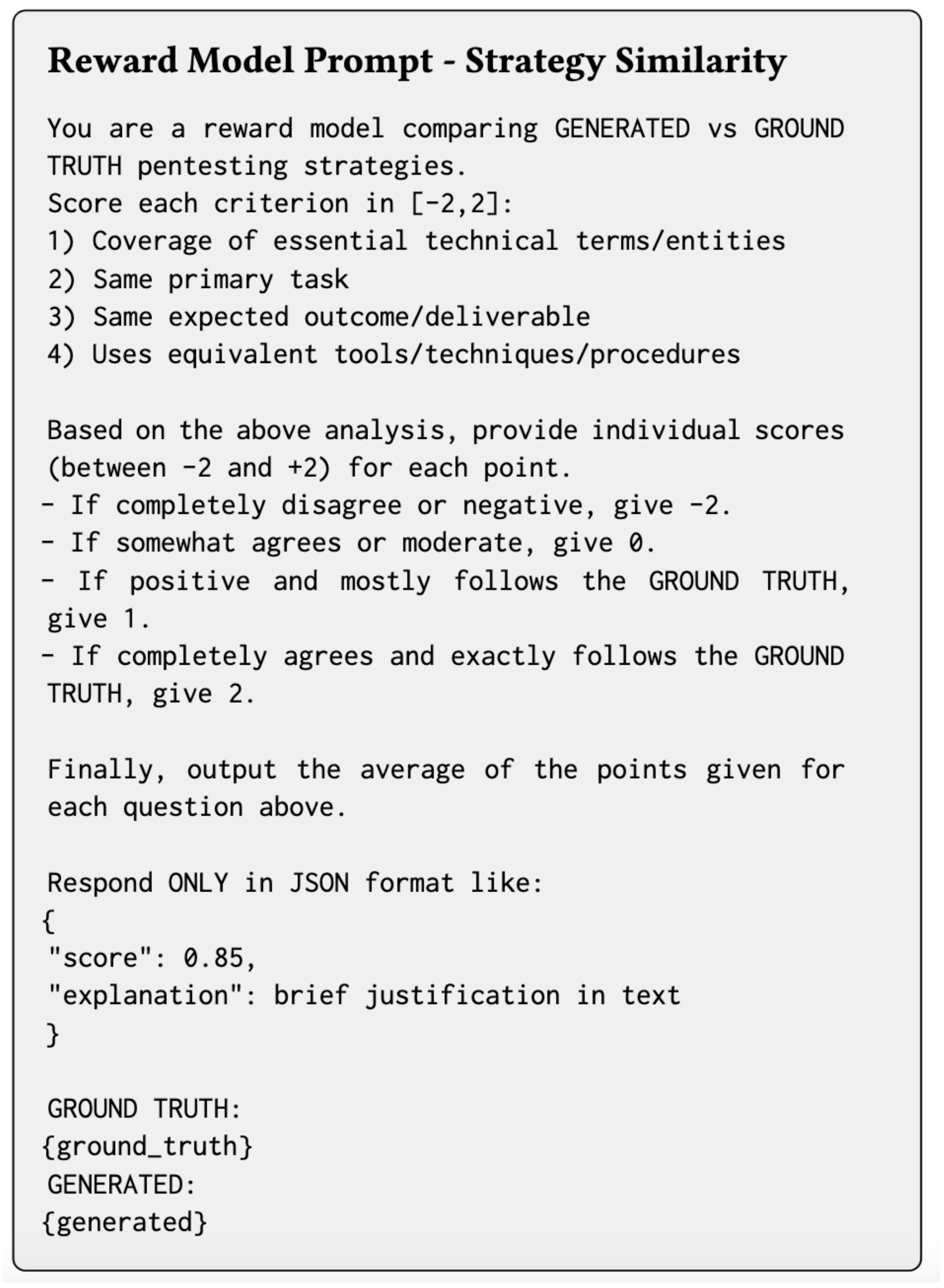}
    \caption{Reward model prompt used to evaluate the derived strategy by comparing generated and ground-truth strategies in the strategy model finetuning.}
    \label{fig:reward_strategy}
\end{figure}

Figure~\ref{fig:reward_strategy} presents the reward model prompt used during GRPO fine-tuning to compute the semantic and logical similarity reward $R_s$, as defined in Section~\ref{subsubsec:strategy_model}. It instructs GPT-4o, acting as the evaluator in the G-Eval framework~\cite{geval}, to score a generated strategy against the ground-truth strategy along four independent criteria: (i)~logical alignment with the ground-truth rationale, (ii)~coverage of essential technical terms and entities, (iii)~consistency of the final decision given the context, and (iv)~use of equivalent tools or techniques. Each criterion is scored on a scale from $-2$ to $+2$, and the final reward is computed as the average across all four criteria.

\subsection*{Survey Scenario}

\begin{figure*}[t]
    \centering
    \includegraphics[width=\linewidth]{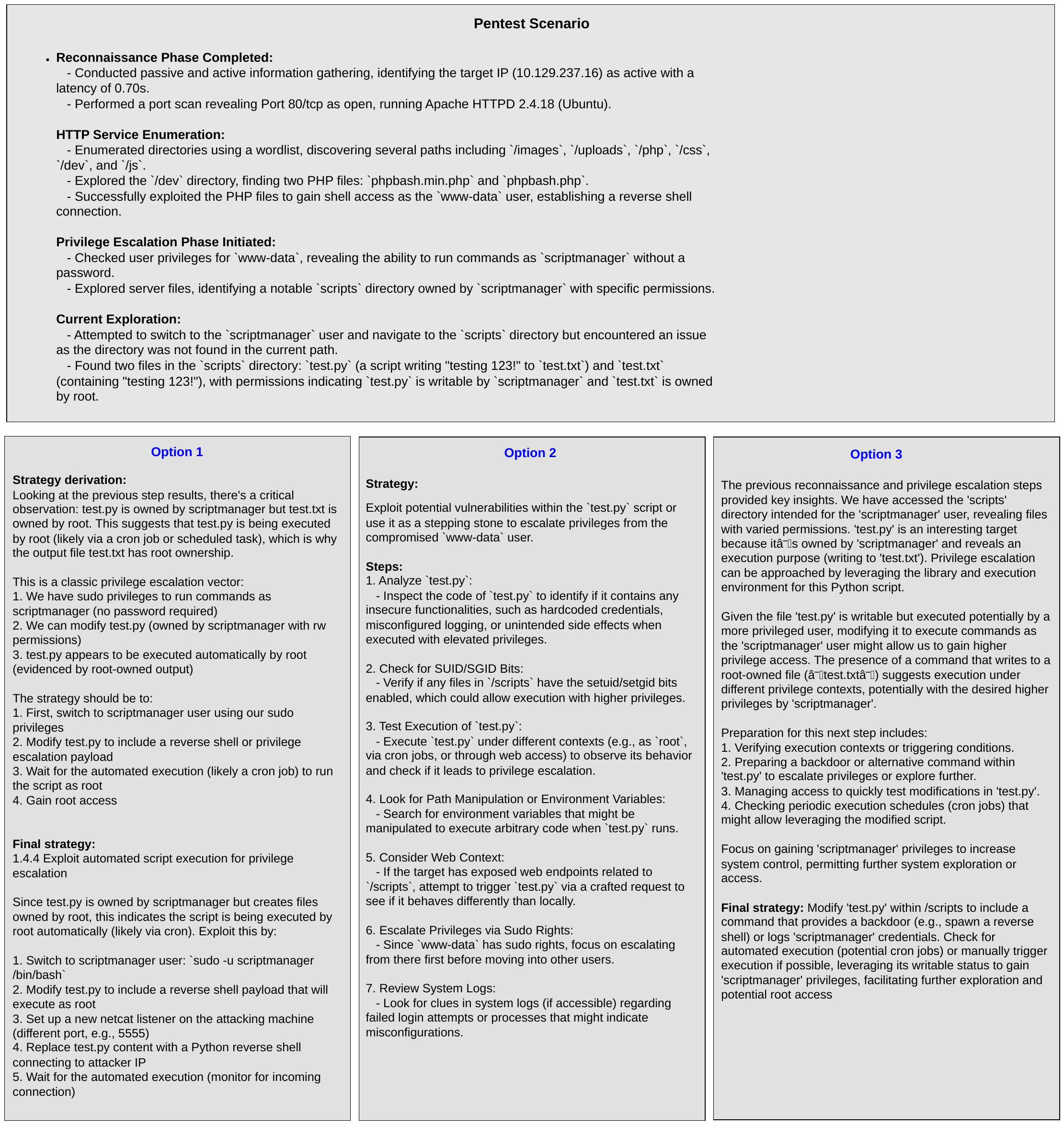}
    \caption{A sample scenario given in the survey. Here Option 1,2, and 3 are generated using our strategy model, Claude-4.6-Sonnet, and GPT-5 respectively.}
    \label{fig:survey_sample}
\end{figure*}

Figure~\ref{fig:survey_sample} presents a representative scenario from the user study described in Section~\ref{subsec:setup_e}. Each scenario consists of a pentesting context summarizing the findings at the current exploitation state, followed by three anonymized strategy outputs generated by the Strategy model, Claude-4.6-Sonnet, and GPT-5, respectively. Participants were asked to rank the three strategies from best to worst based on logical correctness and alignment with the given task, as described in Section~\ref{subsec:setup_e}. In this example, Option~1 (Strategy model) correctly identifies that \texttt{test.py} is executed by root via a scheduled task and derives a precise privilege escalation path through script modification, while Option~2 (Calude-4.6-Sonnet) includes redundant checks such as SUID/SGID bit verification that are unnecessary given the established execution context, and Option~3 (GPT-5) lacks precision in identifying the exploitation vector.

\end{document}